\documentclass[openany]{nature}
\usepackage{graphicx,psfrag}
\usepackage{mathrsfs}
\usepackage{amsmath,amsfonts,amssymb}
\usepackage{multirow}
\usepackage{comment}
\usepackage{units}
\usepackage{enumerate}
\usepackage[normalem]{ulem} 
\usepackage{longtable}
\usepackage[rgb]{xcolor}
\usepackage{gensymb}
\usepackage{adjustbox}
\usepackage[none]{hyphenat}
\usepackage{comment}

\usepackage{xcolor}

\newcommand{\kms}{\,km\,s$^{-1}$}
\newcommand{\mn}{{Mon. Not. R. Astron. Soc.}}
\newcommand{\mnras}{\mn}
\newcommand{\aj}{{"Astron. J."}}
\newcommand{\apj}{{Astrophys. J.}}
\newcommand{\apjl}{{Astrophys. J. Lett.}}

\newcommand{\aap}{{Astron. Astrophys.}}
\newcommand{\nat}{{Nature}}

\newcommand{\pasp}{{Pub. Ast. Soc. Pac.}}

\newcommand\ion[2]{#1$\;${%
\ifx\@currsize\normalsize\small \else
\ifx\@currsize\small\footnotesize \else
\ifx\@currsize\footnotesize\scriptsize \else
\ifx\@currsize\scriptsize\tiny \else
\ifx\@currsize\large\normalsize \else
\ifx\@currsize\Large\large
\fi\fi\fi\fi\fi\fi
\rmfamily\@Roman{#2}}\relax}%
\renewcommand{\ion}[2]{{\rm#1}\;\textsc{\MakeLowercase{#2}}}
\newcommand{\au}{\mathrm{AU}}

\title{High-resolution [\ion{O}{I}] line spectral mapping of TW Hya supportive of a magnetothermal wind}

\author{Min Fang$^{1, 2}$ $^{*}$, Lile Wang$^{3, 4}$, Gregory J. Herczeg$^{3, 4}$, Jun Hashimoto$^{5,6,7}$, Ziyan Xu$^{8}$, Ahmad Nemer$^{9}$, Ilaria Pascucci$^{10}$, Sebastiaan Y. Haffert$^{11}$, \& Yuhiko Aoyama$^{3}$}

\begin{document}

\maketitle

\begin{affiliations}
\item Purple Mountain Observatory, Chinese Academy of Sciences, 10 Yuanhua Road, Nanjing 210023, China
\item University of Science and Technology of China, Hefei 230026, China
\item Kavli Institute for Astronomy and Astrophysics, Peking University, 5 Yiheyuan Road, Beijing 100871, China
\item Department of Astronomy, School of Physics, Peking University, 5 Yiheyuan Road, Beijing 100871, China
\item Astrobiology Center, National Institutes of Natural Sciences, 2-21-1 Osawa, Mitaka, Tokyo 181-8588, Japan
\item Subaru Telescope, National Astronomical Observatory of Japan, Mitaka, Tokyo 181-8588, Japan
\item Department of Astronomy, School of Science, Graduate University for Advanced Studies (SOKENDAI), Mitaka, Tokyo 181-8588, Japan
\item Univ Lyon, Univ Lyon 1, ENS de Lyon, CNRS, Centre de Recherche Astrophysique de Lyon UMR5574, F-69230, Saint-Genis-Laval, France
\item Center for Astro, Particle and Planetary Physics (CAP3), New York University Abu Dhabi, PO Box 129188, Abu Dhabi, UAE
\item Department of Planetary Sciences, University of Arizona, 1629 East University Boulevard, Tucson, AZ 85721, USA
\item Steward Observatory, University of Arizona, 933 North Cherry Avenue, Tucson, Arizona, USA
\end{affiliations}

\begin{abstract}
Disk winds are thought to play a critical role in the evolution and dispersal of protoplanetary disks.  A primary diagnostic of this physics is emission from the wind, especially in the low-velocity component of the [\ion{O}{I}]~$\lambda6300$ line. 
However, the interpretation of the line is usually based on spectroscopy alone, which leads to confusion between magnetohydrodynamic winds and photoevaporative winds. Here, we report that in high-resolution VLT/MUSE spectral mapping of TW~Hya, 80\% of the [\ion{O}{I}] emission is confined to  within 1\,AU radially from the star.  A generic model of a magnetothermal wind produces [\ion{O}{I}] emission at the base of the wind that broadly matches the flux and the observed spatial and spectral profiles. The emission at large radii is much fainter than predicted from models of photoevaporation, perhaps because the magnetothermal wind partially shields the outer disk from energetic radiation from the central star. This result calls into question the previously assessed importance of photoevaporation in disk dispersal predicted by models

\end{abstract}

\clearpage

Descriptions of the physics that governs disk evolution and accretion physics depend on our evaluation of the launching of disk winds versus radius\cite{pascucci22}. A high velocity jet is launched near the star-disk interaction region\cite{frank14}, while at larger radii non-ideal magnetohydrodynamic effects\cite{ferreira06,Bai17} and disk irradiation combine to launch slower winds\cite{wang19}.  The picture for disk winds has been built from empirical evidence in many diagnostics, most especially [\ion{O}{I}] emission\cite{hartigan95,rigliaco13,natta14,simon16}, and from numerical simulations, but tests of disk physics are limited by the challenge of distinguishing contributions from magnetized and photoevaporative winds.

Spatially resolving the low velocity component of the wind  can break the degeneracy between various theoretical wind models, however the small physical scales are beyond the limits of most previous observations. The adaptive optics mode of the MUSE instrument of the Very Large Telescope (VLT) provides a powerful new capability to measure the spatial extent of optical emission in winds.  We further optimize the physical scale by focusing on TW~Hya, the nearest ($\sim$60\,pc, ref.\cite{gaiaedr3})  solar-mass star that is still actively accreting from its protoplanetary disk.  The disk is viewed face-on (disk inclination$\sim5^\circ-7^\circ$, ref.\cite{qi04,hughes11,teague16,huang18}), thereby offering the best possible physical resolution for measuring the radial distribution of emission. High-resolution ALMA observations reveals an inner cavity of 1 AU in the distribution of mm-sized dust grains\cite{andrews16}; a similar cavity in small dust grains is measured from the deficit in near- and mid-IR emission\cite{calvet02}, although some small grains are located near the disk truncation radius\cite{gravity20}. 
Blueshifted [\ion{Ne}{II}] emission from TW~Hya must arise beyond the dust cavity and may indicate the presence of a photoevaporative flow\cite{pascucci09}. The [\ion{O}{I}]\,$\lambda6300$ emission line from TW~Hya is narrow (full width of half maximum of $\sim$13.0\,\kms) with a small blue-shift\cite{pascucci11,simon16,fang18,banzatti19}, measured here with a mean of $-0.8$\,\kms\ (see Extended Data Figures~\ref{Fig:OIcen} and~\ref{Fig:lineprofile}. The spectroscopic comparison with blueshifted [\ion{Ne}{II}] emission leads to the inference that the [\ion{O}{I}] emission is likely more compact and may even originate in the disk and not a wind\cite{pascucci11,pascucci20}.

\noindent{\large\bf Results} 

\begin{figure}
\includegraphics[width=3.25in]{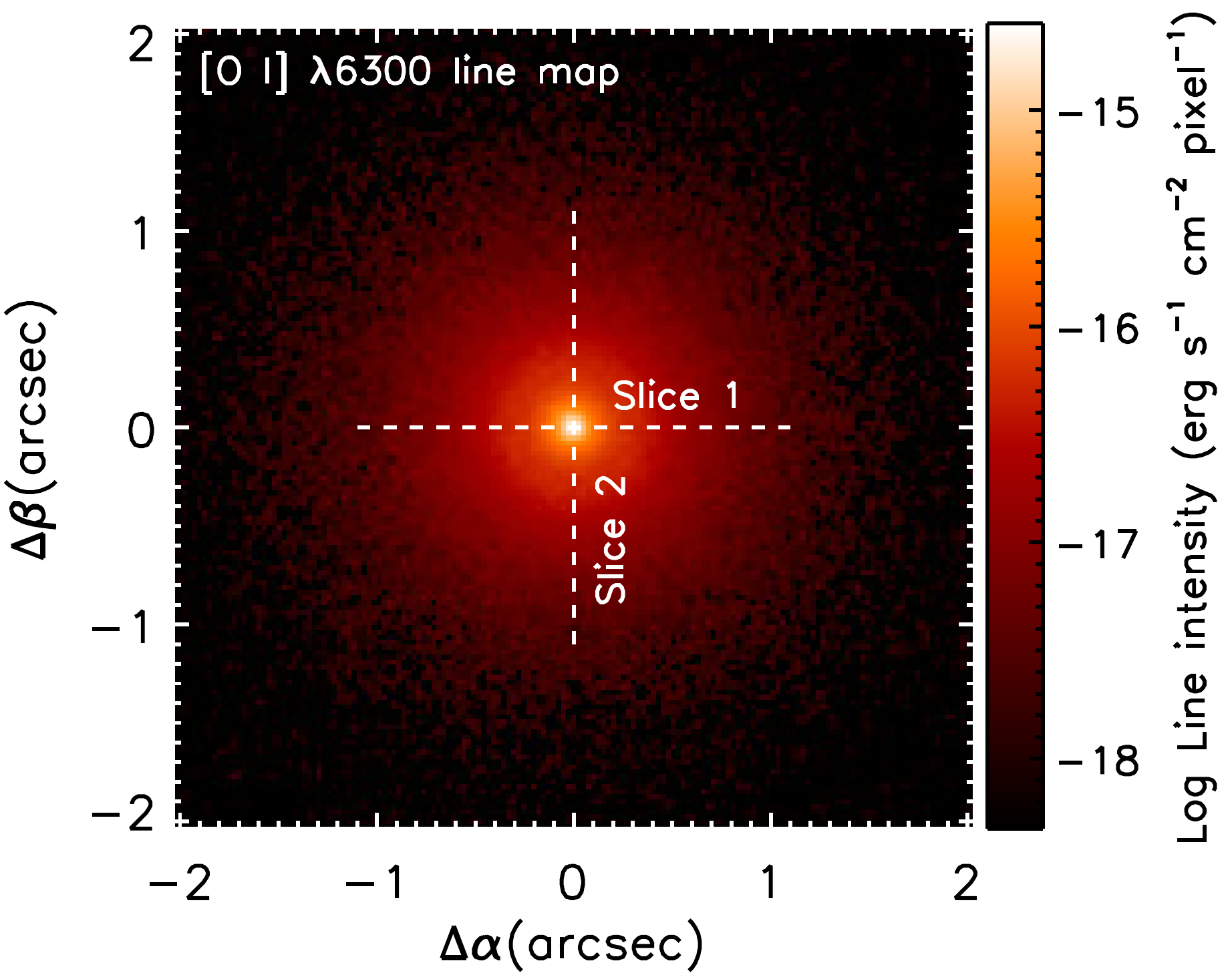}
\includegraphics[width=3.25in]{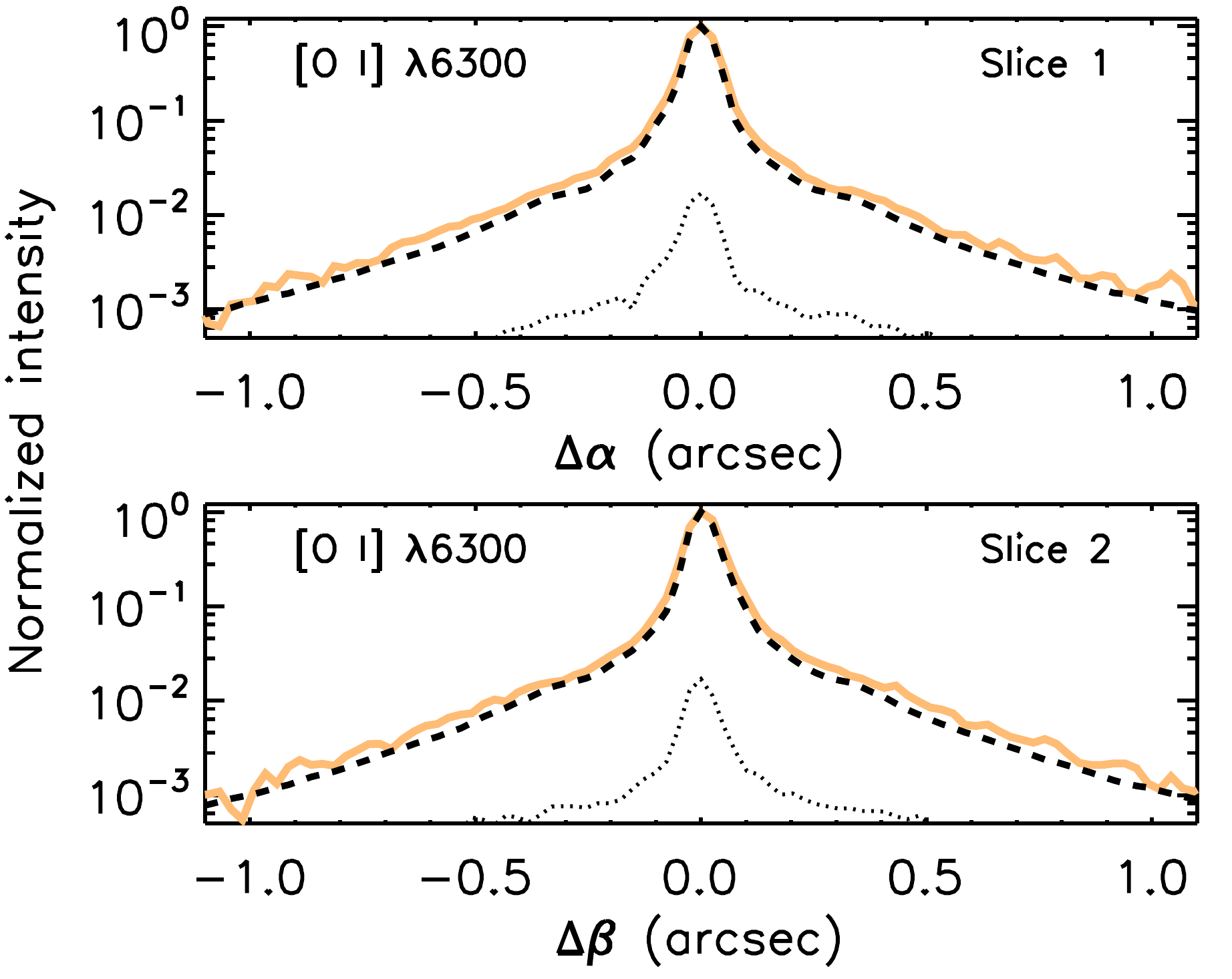}
\includegraphics[width=3.25in]{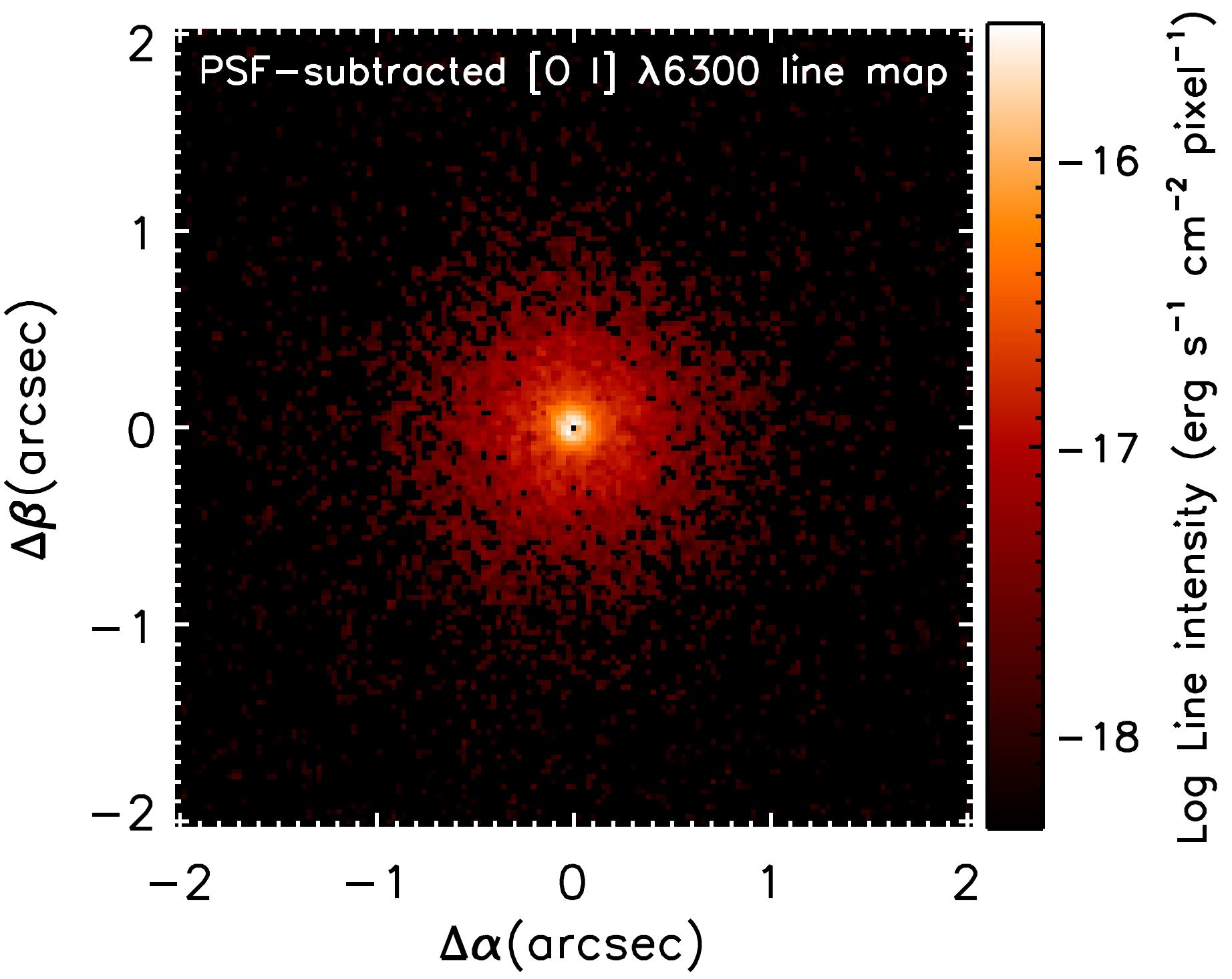}
\includegraphics[width=3.25in]{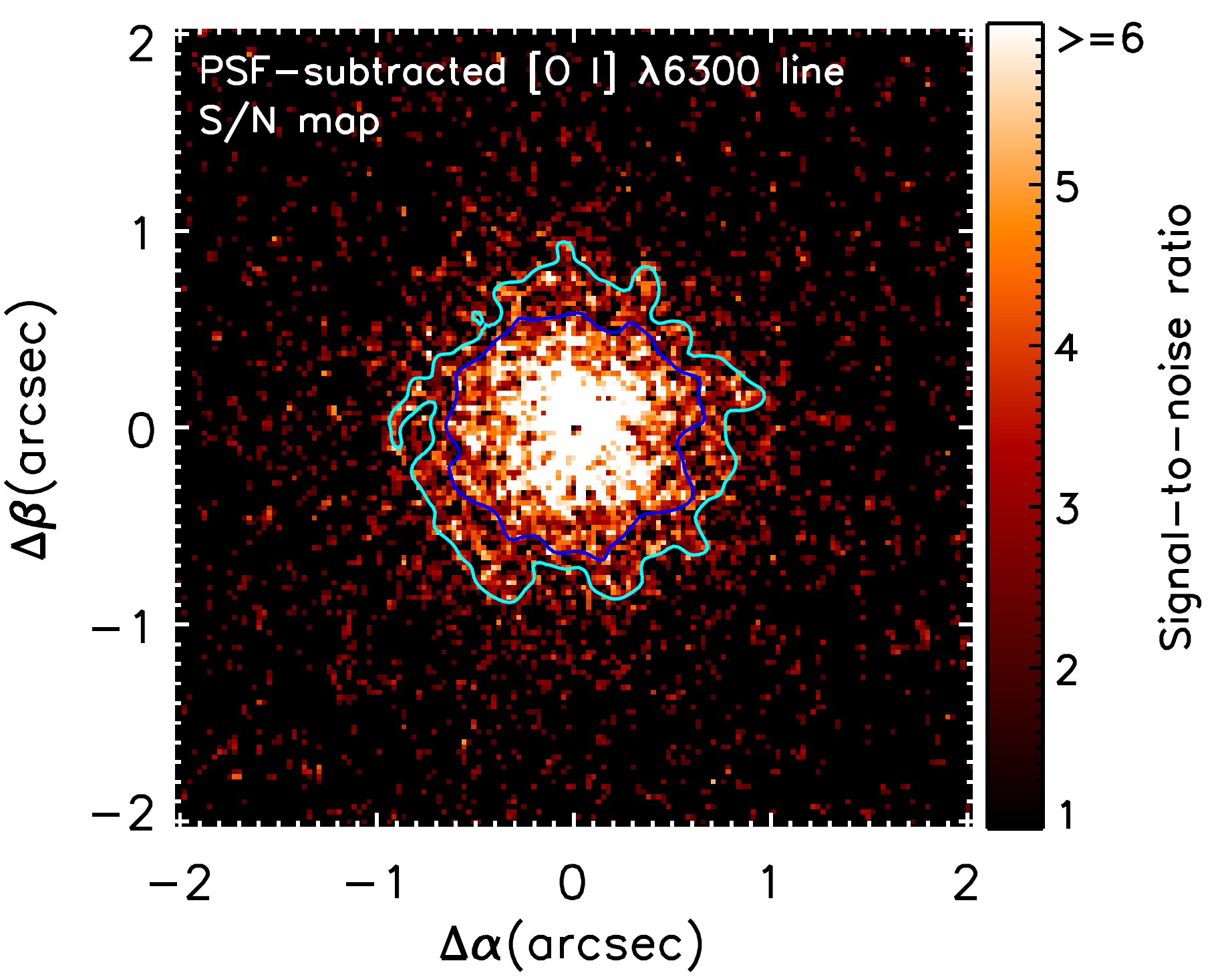}
\linespread{1.0}\selectfont{}
\caption{\textbf{Error-weighted mean [\ion{O}{I}]~$\lambda6300$  intensity map.} Upper-left panel: Error-weighted mean [\ion{O}{I}]~$\lambda6300$ intensity map of TW~Hya. Upper-right panels: Two slice cuts (yellow lines) for  [\ion{O}{I}]~$\lambda6300$ line map shown in the upper-left panel, overplotted with the PSF (dark dashed lines) near [\ion{O}{I}]~$\lambda6300$ line. The dotted line in each panel is the standard deviation along the slice cut. Lower-left panel: [\ion{O}{I}]~$\lambda6300$ line residual map after a subtraction of a PSF which has been normalized to line map by the peak emission.  Lower-right panel: the signal-to-noise ratio map for the [\ion{O}{I}]~$\lambda6300$ line residual map shown in the Lower-left panel. The cyan and blue lines show contours with S/N=2 and 3, respectively. The contours have been smoothed with a Gaussian function with a full width of half maximum of 5 pixels.}
\label{Fig:OI6300_mean}
\end{figure}

Figure~\ref{Fig:OI6300_mean} (upper-left panel) shows the error-weighted mean [\ion{O}{I}]~$\lambda6300$ intensity map of TW~Hya with MUSE  (See Methods for a detail description and Extended Data Figure~\ref{fig:example_line_OI6300_HeI6678} for [\ion{O}{I}] line emission detected with MUSE). When compared with the continuum emission, vertical and horizontal slices (upper right panels) show enhanced  [\ion{O}{I}] emission in both directions. The continuum point spread function (PSF) is scaled to this emission and subtracted, leaving a residual signal detected out to $\sim1''$ (60\,AU), as shown in the intensity map and signal-to-noise (S/N) ratio map for the PSF-subtracted [\ion{O}{I}]~$\lambda6300$ emission in the bottom panels of Figure~\ref{Fig:OI6300_mean}. To further establish the significance of this detection, we follow the same procedures for the \ion{He}{I}~$\lambda6678$  emission line, which is produced by accretion processes close to the star and is not expected to be spatially extended\cite{yang07}.  The \ion{He}{I}~$\lambda6678$  emission is as compact as the nearby continuum emission (see Extended Data Figure~\ref{Fig:HeI6678_mean}). The comparison with  \ion{He}{I} emission, in combination with the signal-to-noise ratio map for the [\ion{O}{I}]~$\lambda6300$ intensity residual map, indicates that the extended emission from [\ion{O}{I}]~$\lambda6300$ is substantial.

\begin{figure}
\includegraphics[width=3.25in]{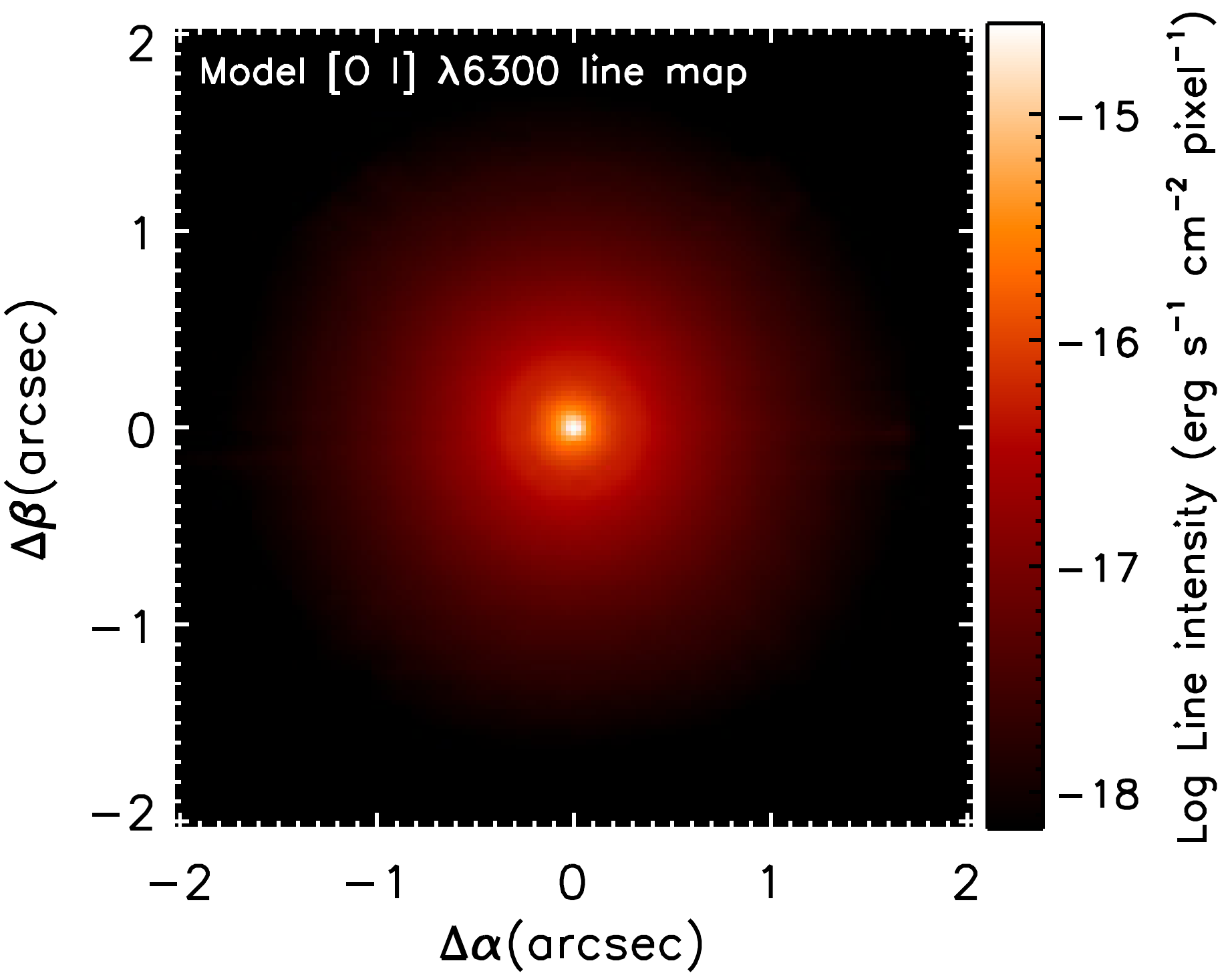}
\includegraphics[width=3.25in]{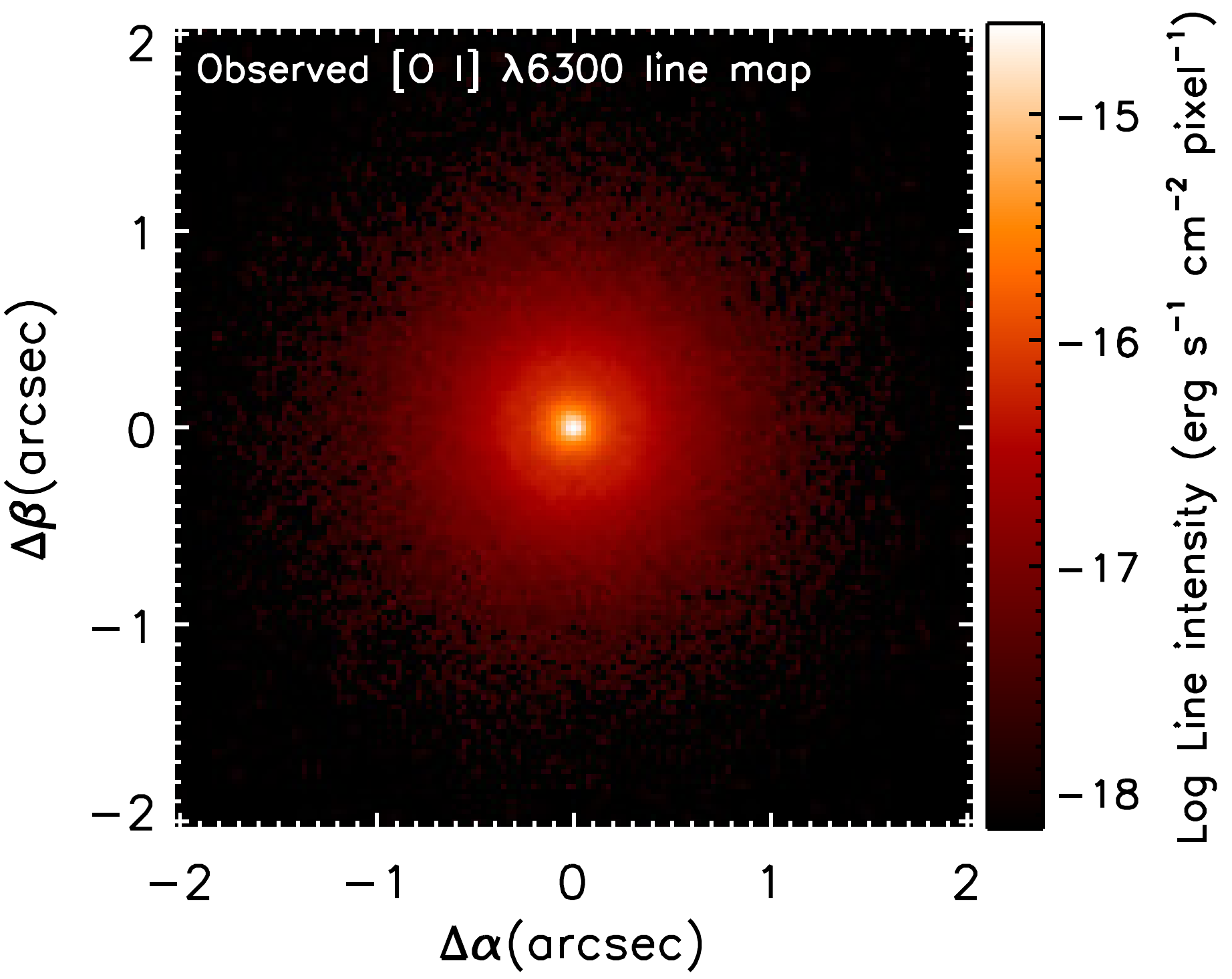}
\includegraphics[width=3.25in]{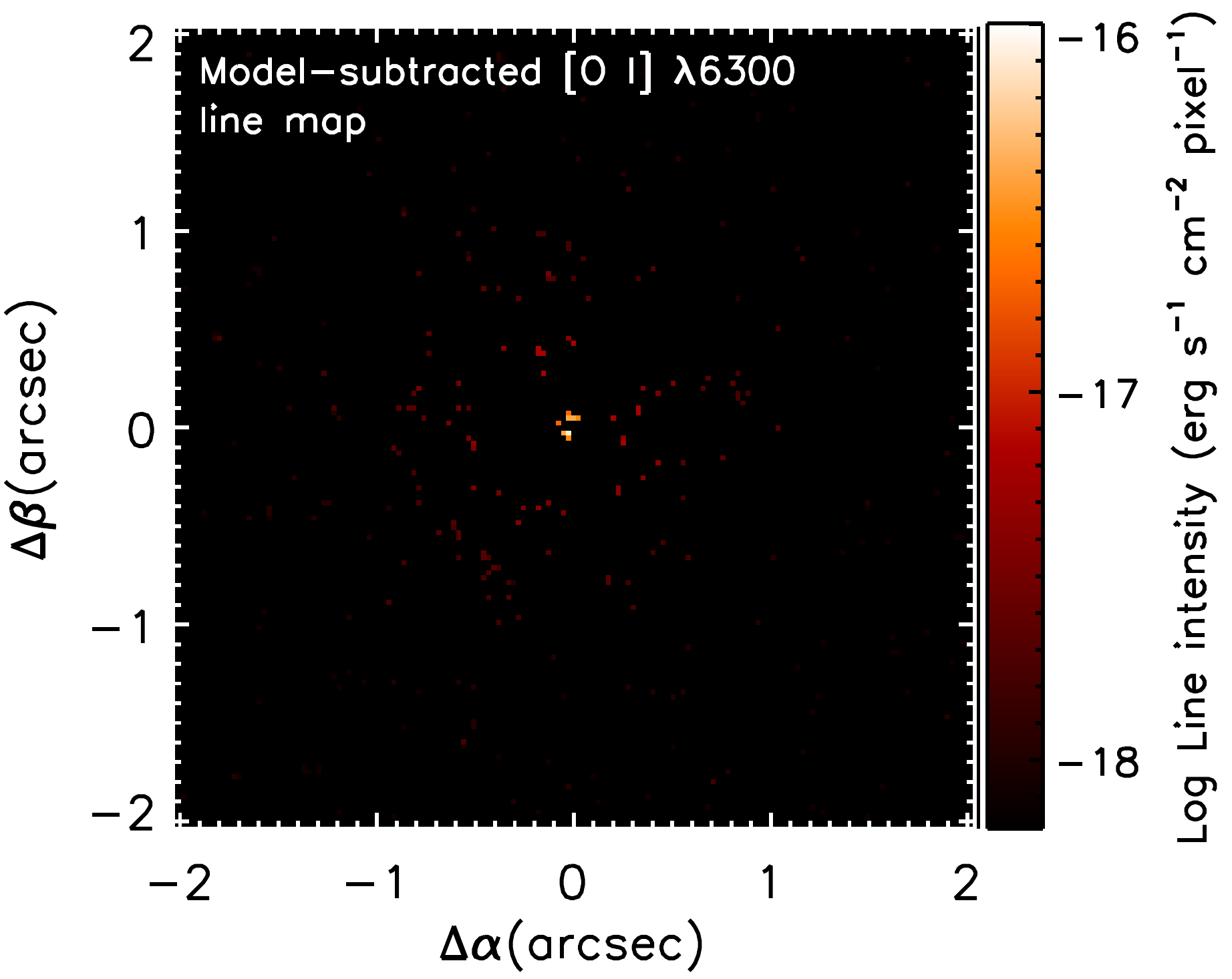}
\includegraphics[width=3.25in]{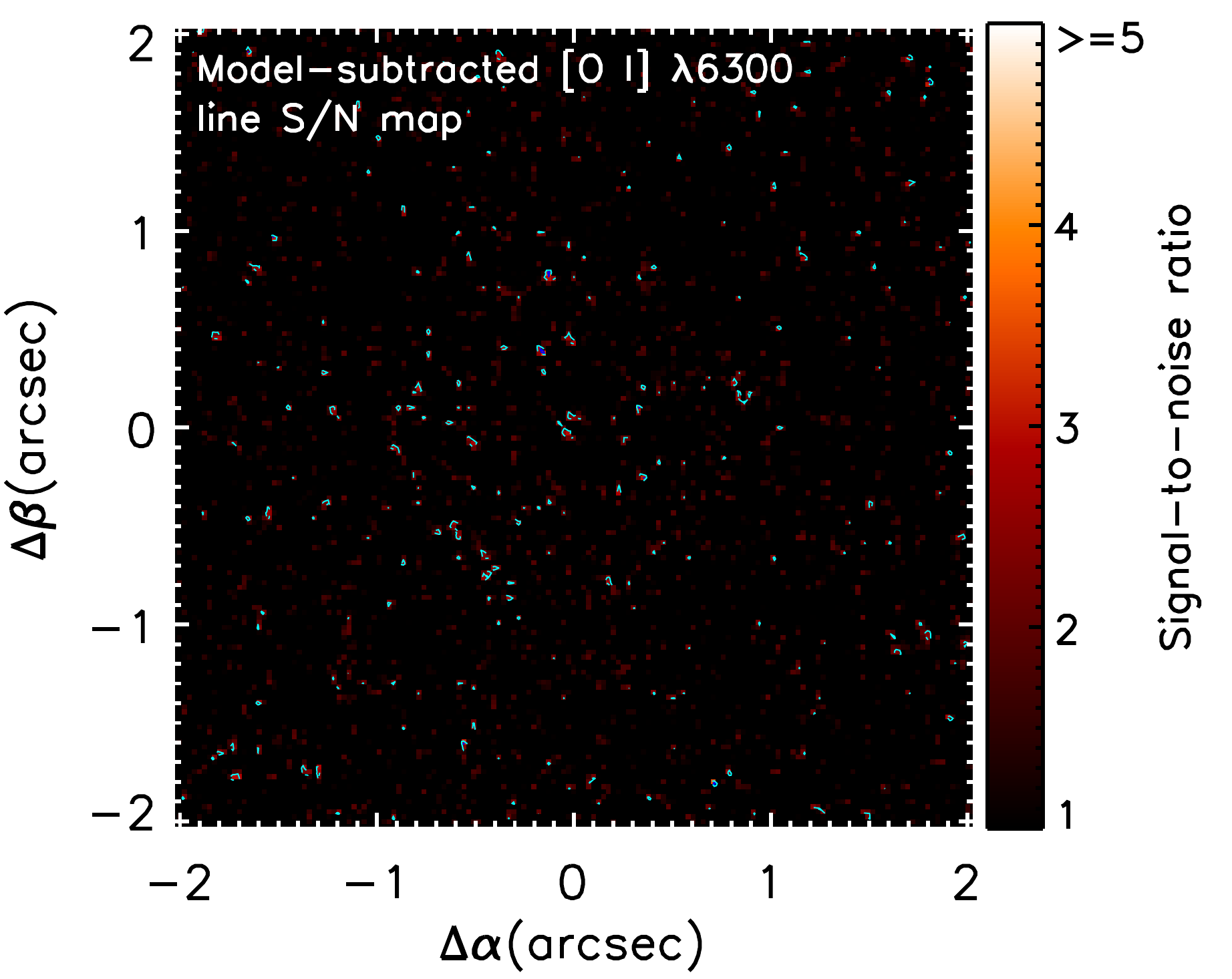}
\linespread{1.0}\selectfont{}
\caption{\textbf{Best-fit model of  [\ion{O}{I}]~$\lambda6300$  intensity map.} Upper-left panels: the best-fit model [\ion{O}{I}]  intensity map. Upper-right panel: error-weighted mean [\ion{O}{I}] intensity map of TW~Hya.  Lower-left panel: [\ion{O}{I}] residual map after a subtraction of the best-fit model. Lower-right panel: the signal-to-noise ratio map for the [\ion{O}{I}]  residual map shown in the Lower-left panel. The cyan and blue lines show contours with S/N=2 and 3, respectively.}\label{Fig:OI6300_model}
\end{figure}

 To evaluate the spatial distribution of [\ion{O}{I}] emission, we create a toy model with a power-law radial profile for the [\ion{O}{I}] intensity, $I_{6300}\propto R^{-\alpha}$, where $R$ is the disk radius. The best-fit model (See Methods for a detail description and Extended Data Figure~\ref{Fig:reduce_Chi2} for the reduced $\chi^2$) is shown in  Figure~\ref{Fig:OI6300_model}, with an inner radius ($R_{\rm in}$) of $\sim$0.08\,AU, a radial power law of $\alpha=2.6$,  and a poorly constrained outer radius of $>40$\,AU.  About 80\% of emission in this model is emitted within 1\,AU, confirming previous inferences from line profile fitting alone\cite{pascucci20}.  The intensity map for the best-fit model accurately reproduces the observed intensity map, with no significant residuals (see Figure~\ref{Fig:OI6300_model}).

\noindent{\large\bf Discussion}

The compact spatial distribution of [\ion{O}{I}] provides strong constraints on the two categories of wind launching mechanisms from the disk, namely (1) photoevaporative winds, and (2)
magnetothermal winds, discussed in turn below.  Numerical studies, starting with simplified disk hydrodynamics and thermochemistry models\cite{Alexander06a, Alexander06b, Gorti08, Gorti09}, have progressively developed the theoretical framework of both mechanisms, with predictions of different spatial profiles of forbidden line emission, including [\ion{O}{I}], that could be used to distinguish between them\cite{nemer20}.

Photoevaporative winds are launched by injecting thermal energy through photoionization and photodissociation, a process that requires the thermal energy in heated gas being greater than the depth of the local potential well.  This requires the wind launching point to be located outside a critical radius, which roughly equates the local effective potential well depth to the specific thermal energy\cite{Alexander06b}. 
In photoevaporative wind models driven by soft X-rays\cite{Owen10,Owen12,Ercolano2016}, the outflows are weak within $\sim$2\,AU under plausible irradiation luminosities. Extreme ultraviolet (EUV) driven photoevaporation applies to regions with
 higher temperatures and smaller mean molecular mass\cite{Wang17},  but is still unable to launch winds efficiently within $R\lesssim 1$~AU of the star.  As a result, the innermost $\sim 1$~AU does not contribute substantially to photoevaporative flows 
 due to the excessive potential well depth.

 \begin{figure*}
\begin{center}
\includegraphics[width=6.2in]{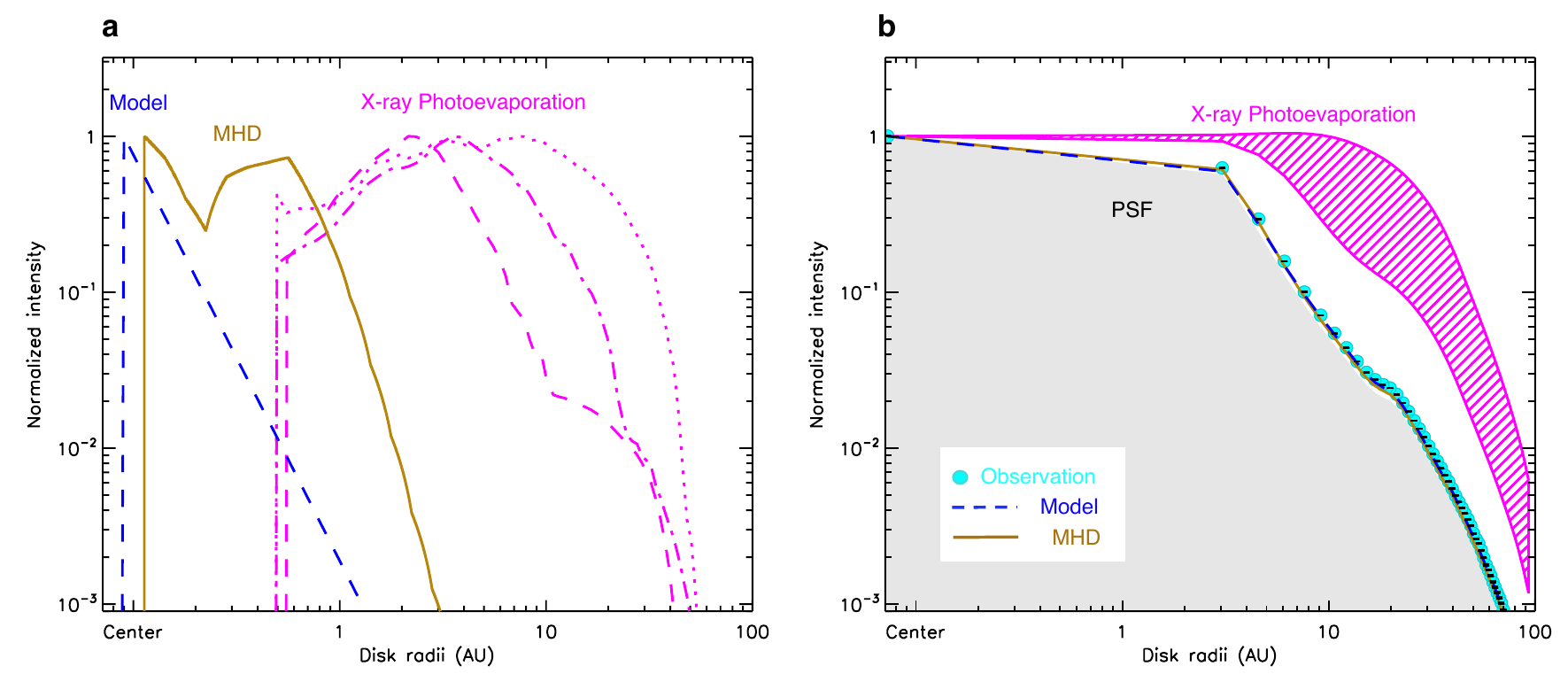}
\linespread{1.0}\selectfont{}
\caption{\textbf{Photoevaporation and magnetothermal models of [\ion{O}{I}]~$\lambda6300$ map.} Left: the best-fit model (blue dashed line) compared with the predictions from X-ray photoevaporation (magenta dotted, dash-dotted, and dashed lines for $\log~L_{\rm X}$~(erg~s$^{-1}$)=28.3, 29.3, and 30.3, respectively)\cite{Ercolano2010}, and from the MHD wind model (yellow solid line). Each model is normalized by its peak intensity. Right: the radial profile  (light blue color filled circles) of the observed [\ion{O}{I}]  line compared with the PSF-convolved best-fit model (blue dashed line), the PSF (gray color filled area), the predictions, convolved with the PSF, from X-ray photoevaporation (the magenta line-filled area which covers the predictions with $\log~L_{\rm X}$~(erg~s$^{-1}$)=28.3$-$30.3, ref.\cite{Ercolano2010}) and MHD wind model. Each model is normalized by its central intensity. The colors are same as in the left panel.  \label{Fig:PEmodel}}
\end{center}
\end{figure*}

The left panel of Figure~\ref{Fig:PEmodel} compares the best-fit model from the [\ion{O}{I}] intensity map of TW~Hya with the predictions from X-ray photoevaporation for X-ray luminosities ($L_{\rm X}$) with $\log~L_{\rm X}$~(erg~s$^{-1}$)=28.3$-$30.3 (ref.\cite{Ercolano2010}), with the bright end similar to the measured X-ray luminosity of TW~Hya of $\log~L_{\rm X}$~(erg~s$^{-1}$)=30.2 (ref.\cite{Kastner02,Stelzer04}).
In the right panel, we convolved the models with the PSF of the MUSE observations near the  [\ion{O}{I}] line and compare them with the  observed  [\ion{O}{I}]  intensity map. 
 Even using models with a smaller inner disk radius (the dust sublimation radius), the X-ray photoevaporation models overpredict the extensions of  [\ion{O}{I}] line emission. An increase of $R_{\rm in}$ will extend the [\ion{O}{I}] intensity map and decrease the  [\ion{O}{I}] flux from inner disk radii\cite{Ercolano2010}.

Magnetothermal winds (launched by magnetic pressure gradients\cite{Bai16} rather than the centrifugal mechanism\cite{blandford82}) are, in contrast, not inhibited in the inner zones. Their wind speeds ($v_{\rm wind}$) are roughly proportional to the local Keplerian velocities (hence
$v_{\rm wind}\propto R^{-1/2}$) -- the closer the wind base is located to the star, the faster the wind is launched\cite{Bai13, Bai17, Wang17, Wang19}.  Assuming that the differential wind mass-loss rates ${\rm d} \dot{M}/{\rm d}\ln R$ are roughly uniform (likely caused by the re-distribution of magnetic fluxes inside the disk\cite{Bai17b,Wang19}), the density of winds ($\rho_{\rm wind}$) near the wind bases roughly scales as
\begin{equation}
  \label{eq:scale-wind-density}
  \rho_{\rm wind}\propto \dfrac{{\rm d}\dot{M}}
  {{\rm d}\ln R} v_{\rm wind}^{-1} R^{-2}\propto R^{-3/2}\ .
\end{equation}
If we further assume that the fractions of \ion{O}{I} and free electrons are mostly invariant inside the magnetothermal winds, the differential emission power would roughly scale as ${\rm d}L_{[\ion{O}{I}]}/{\rm d}R\propto R^{-1}$, indicating that most of the [\ion{O}{I}] emission concentrates near the innermost wind-launching region. 

Figure~\ref{Fig:PEmodel} also compares the [\ion{O}{I}] intensity map of TW~Hya with the prediction from magnetothermal winds based on the non-ideal magnetohydrodynamic (MHD) simulations with consistent
thermochemistry\cite{Wang19}, and the subsequent far-ultraviolet pumping emission line synthesis\cite{nemer20}. This generic magnetothermal wind model has a specific wind mass-loss rate of  ${\rm d}\dot{M}_{\rm wind}/{\rm d}\ln R \sim 3\times10^{-9}$~M$_\odot$ yr$^{-1}$ within 2 AU. The generic model also uses a full disk and not an inner hole, which will affect emission from the back side of the disk.
 Without fine-tuning, this model predicts a spatial and spectral profile that largely matches the observations of [\ion{O}{I}] (see a comparison in Figure~\ref{Fig:PEmodel} and Extended Data Figure~\ref{Fig:lineprofile}).  The model produces $\sim 1.2\times10^{-5}$ L$_\odot$ in the [\ion{O}{I}] line, which is consistent with the observed luminosity of $\sim 1.5\times10^{-5}$ L$_\odot$.  The hole in the inner disk allows some of emission on the back side of the disk to be detected, which will decrease the centroid velocity from our generic full disk model\cite{pascucci11,pascucci20}.

 The [\ion{Ne}{II}] emission line should trace different spatial regions in the winds, as it peaks at slightly outer radii ($\sim 1~\au$) and higher altitudes in the magnetothermal wind models. Nonetheless, the model [\ion{Ne}{II}] emission exhibits strong agreement with the observed line profile\cite{pascucci11} with a luminosity three times weaker than measured\cite{pascucci09} (see Extended Data Figure~\ref{Fig:lineprofile}).

The spatially compact emission region is explained by the magnetothermal winds rather than photoevaporation. The synthesized emission from photoevaporative models\cite{Ercolano2010} expands the major emission regions up to $\sim25~\au$.  Detailed analyses based on self-consistent magnetothermal wind models, including non-LTE effects including the pumping by FUV photons, confirm that the spatial regions of [\ion{O}{I}] emission are concentrated within the innermost $\sim$1\,AU(ref.\cite{nemer20}). 
Photoevaporation may add to the mass loss from high altitudes of the innermost disk, but only when magnetothermal winds are already blowing\cite{Wang19}. Moreover, since the magnetothermal winds enhance the gas density in the disk coronal regions, many photons (including EUV and soft X-ray) from the star that drive photoevaporation  should not be able to reach disk surfaces due to effective shading. Therefore, the comparison with theoretical models guides us to the conclusion that the magnetothermal wind is the indispensable mechanism behind the spatial compactness of [\ion{O}{I}] emission zones.

The MUSE observations definitively place most of the [\ion{O}{I}] emission from the inner 1\,AU of the disk.  Magnetothermal winds recover both the spatial compactness, the flux, and the narrow, slightly blueshifted line profile for [\ion{O}{I}] emission from the base of the disk wind.  
These conclusions provide us with a new perspective to re-evaluate decades of high spectral resolution observations of [\ion{O}{I}] emission\cite{hartigan95,rigliaco13,simon16,banzatti19}.  The emission at large disk radii is  more than one order of magnitude fainter than expected from photoevaporation models, which may indicate that photoevaporation rates are less than expected; photodissocation of H$_2$O and scattering of vertically elevated [\ion{O}{I}] emission may also contribute to this spatially extended component.
In the magnetothermal wind models the weak photoevaporation at large radii results from opacity in the wind from the inner disk shielding the outer disk from high energy photons\cite{pascucci22}. Reducing irradiation of the outer disk would decrease the contribution of photoevaporation in disk dispersal and would also alter the chemistry by decreasing ionization and photodissociation rates.   The blueshifted [\ion{Ne}{II}] emission has been interpreted as evidence for photoevaporation, however a similar spectral line profile is produced with the generic magnetothermal wind.  A similar spatial analysis as performed here is needed to measure the relative contributions to the production of [\ion{Ne}{II}] emission.

These advances are possible because of the high angular resolution offered by MUSE.  Previous observations of winds at such high angular resolution, including with MUSE, have focused mostly on collimated 
jets\cite{lavalley97,schneider13,schneider20,xie21,flores-rivera23}.  Prior spectro-astrometry of [\ion{O}{I}] emission from a different disk, RU~Lup, indicates a vertical extent of at least 10~AU, without the 2D imaging required to describe the spatial distribution of the emission in the disk plane\cite{whelan21}.  For TW~Hya, the observational results confirm previous inferences from spectroscopy that the [\ion{O}{I}] emission is produced within a compact region\cite{pascucci20}. Our empirical discovery that the  [\ion{O}{I}] emission is produced within $\sim 1 AU$ is then interpreted {as emission from the base of the magnetothermal disk wind, based on the a line profile and flux that is consistent} with predictions from models of the wind.  The empirical measurement is also consistent with some of the [\ion{O}{I}] emission produced in the disk surface, below the wind.

Further observations with MUSE, combined with high resolution spectroscopy and wind models, will help to assess whether the implications for the MHD wind are generalizable. [\ion{O}{I}] line profiles are rarely double-peaked\cite{banzatti19,gangi20}, as expected if a non-turbulent wind from a system like TW Hya were viewed closer to edge-on.  
Some systems that appear young, including RU~Lup, have different characteristic line emission, including [\ion{S}{II}] $\lambda$6716,6731, which with further modeling and observations will allow us to better understand the connection between system properties (age, disk structure, dust composition) and line shapes.

\clearpage


\newpage

\begin{methods}

\subsection{Observations and Data Reduction}

Observations of TW~Hya were obtained with the optical integral field spectrograph MUSE\cite{bacon17} in the Narrow-Field Mode (NFM) in Program 0104.C-0449 (PI Jos De Boer) to search for H$\alpha$ emission from any accreting protoplanets that might be present.
The adaptive optics (AO) system for MUSE consists of the four laser guide star (LGS) facility and the deformable secondary mirror (DSM). The LGS wavefront sensors measure the turbulence in atmosphere, and the DSM corrects a wavefront to provide near-diffraction-limited images. The MUSE field of view in NFM mode covers  7.$''$42~$\times$~7.$''$43 with a spatial sampling of 25~$\times$~25~mas$^{2}$.  

TW~Hya was observed for a total of 2720~s, split into 16 integrations of 160~s and 4 integrations of 40~s.  Individual raw frames are calibrated with the MUSE pipeline v2.8.5 (ref.\cite{weilbacher20}) in two stages. The first stage consists of basic instrumental calibrations (bias and dark subtraction, flat-fielding, wavelength calibration, measurement of line spread function, geometric calibration, and illumination correction). The second stage consists of post-calibrations (flux calibration, sky subtraction, and distortion correction). The absolute flux is calibrated by both an atmospheric extinction curve at Cerro Paranal and a spectro-photometric standard star stored as MUSE master calibrations. The instrumental distortion is corrected by a multi-pinhole mask.  The astrometric precision is calibrated and monitored by observing stellar cluster fields with high astrometric quality from Hubble Space Telescope imaging.  We adopted
the intrinsic wavelength calibration accuracy of $\sim 0.4$\,\AA\ (according to the MUSE user manual) because the sky lines were too faint for any improvement.

The final spectra cover from 4650 to 9300~\AA at a spectral resolution of $\sim2800$ near 6300~\AA.  A gap in the spectrum 5780--6050 \AA\ is caused by the sodium laser.  The point spread function (PSF) at 6300 \AA\ has a Gaussian core with a full width of half maximum (FWHM) of $\sim 0.''061$ and broad wings that extend to $\sim 1.''75$.  The beam encircles 40\% of the flux within $0.''15$ and 90\% within $0.''9$.

\subsection{Line extraction}

We extract the MUSE spectrum of TW~Hya by performing aperture photometry within an aperture of 6 pixels on each wavelength plane of the reduced data cubes of individual exposures. Examples of the extracted spectra near  [\ion{O}{I}]~$\lambda6300$ or \ion{He}{I}~$\lambda6678$ of two exposures are shown in Extended Data Figure~\ref{fig:example_line_OI6300_HeI6678}. After visually inspecting the spectrum near [\ion{O}{I}]~$\lambda6300$ and \ion{He}{I}~$\lambda6678$ for each exposure, we combine eleven exposures for our final line maps and discard nine exposures because of bad pixels near the core of the [\ion{O}{I}]~$\lambda6300$ or \ion{He}{I}~$\lambda6678$ emission.

For [\ion{O}{I}]~$\lambda6300$ line extraction, the photospheric absorption features from 6240 to 6355\,\AA\ are fit with a veiled template spectrum, TWA 14, an M0.5 (ref.\cite{fang21}) star that provides a good match to the spectrum of TW~Hya (M0.5, ref.\cite{herczeg14}). The template spectrum is obtained from the X-Shooter template library\cite{manara13} and degraded to the spectral resolution of MUSE, $R=2800$ at 6300 \AA.  The template is shifted in velocity and veiled by an excess accretion continuum flux,  parameterized as $r_{6320}=\frac{F_{\rm excess,6320}}{F_{\rm phot,6320}}$, where  $r_{6320}$ is the veiling at 6320\,\AA, $F_{excess,6320}$
is the excess flux at 6320\,\AA, and $F_{phot,6320}$ is the photospheric emission at 6320\,\AA. Best fit parameters of $r_{6320}$ and velocity are found by minimizing $\chi^2$ between the veiled template and the TW~Hya MUSE spectrum within two wavelength ranges, 6287--6246\,\AA\ and 6326--6349\,\AA. 
In order to account for slightly different spectral slopes between the TW~Hya and TWA~14 spectra, which were observed with different instruments, a two-order polynomial function is used to fit the ratio between the MUSE spectra and template and then used to correct the template before $\chi^2$ is calculated. Extended Data Figure~\ref{fig:example_line_OI6300_HeI6678} shows the best-fit templates (red dashed lines) for the MUSE spectra of two exposures. For the eleven individual exposures, the best-fit $r_{6320}$ values range from 0.60--0.67, and velocities range from 19--26\,\kms.

We note that a hump feature within 6287--6296\,\AA\ (Extended Data Figure~\ref{fig:example_line_OI6300_HeI6678}) is present and similar for the exposures with the same rotation angle. In order to mitigate the effect on the line extraction from this feature, we average the residual spectra produced by subtracting the MUSE spectra by the best-fit templates for the exposures with the same instrumental setting. Examples of best-fit templates plus the mean residual spectra (blue dash-dotted lines) are shown in Extended Data Figure~\ref{fig:example_line_OI6300_HeI6678}. 

For  each exposure, we match both the best-fit template and residual-corrected  best-fit template to the MUSE spectrum of TW~Hya spaxel-by-spaxel by scaling the flux of template to the MUSE spectrum at 6320\,\AA,  adjusting the difference on the spectral slope between the template and the MUSE spectrum in each spaxel using a second-order polynomial function. In this way, we obtain the best-fitted templates for each spaxel.  We derive the  [\ion{O}{I}] $\lambda6300$ line emission in every spaxel by subtracting the residual-corrected template from the MUSE spectrum.  The [\ion{O}{I}] $\lambda6300$  intensity map for each exposure is then obtained by integrating the [\ion{O}{I}] $\lambda6300$  line emission in each spaxel with an uncertainty estimated using the standard deviation from the template-subtracted residual spectra (see wavelength ranges for the line extraction and the uncertainty estimate in Extended Data Figure~\ref{fig:example_line_OI6300_HeI6678}). The templates used here are not corrected for the instrumental residual in order to avoid underestimating the noise.
To obtain the PSF near the [\ion{O}{I}]~$\lambda6300$ line, we integrate the flux within the wavelength ranges, 6263--6294\,\AA\ and 6308--6339\,\AA\ in each spaxel.  Uncertainties are also estimated using the standard deviation described above.  
 A similar procedure is followed for \ion{He}{I} $\lambda6678$ (see Extended Data Figure~\ref{fig:example_line_OI6300_HeI6678}).  The spectral template is fit to the TW~Hya spectrum from 6620 to 6735~\AA, with best-fit veiling values at 6729\,\AA, $r_{6729}$, ranging from 0.48--0.59, and $v$ values ranging from 19--26\,\kms\ for the used eleven individual exposures. The wavelength ranges for  
 obtaining the PSF near \ion{He}{I}~$\lambda6678$ and estimating the uncertainty are  6636--6668~\AA\ and  6689--6721~\AA.

\subsection{Line maps}

For each exposure, the  [\ion{O}{I}]~$\lambda6300$ emission is extracted from the spectrum in each spaxel as described above. To further establish the significance of this detection, we follow the same procedures to extract the \ion{He}{I}\,$\lambda6678$ intensity map of each exposure. The [\ion{O}{I}], \ion{He}{I}, and their nearby continuum images are then coadded across the eleven exposures used  in our analysis. Figure~\ref{Fig:OI6300_mean} shows the error-weighted mean [\ion{O}{I}]\,$\lambda6300$ intensity map, the PSF along horizontal and vertical slices, the PSF-subtracted [\ion{O}{I}]\,$\lambda6300$ intensity map, and the signal-to-noise ratio map for the PSF-subtracted [\ion{O}{I}]\,$\lambda6300$ intensity map. Extended Data Figure~\ref{Fig:HeI6678_mean} provides those same maps for \ion{He}{I}\,$\lambda6678$ line emission. A comparison of the error-weighted mean [\ion{O}{I}] and \ion{He}{I} images with their nearby continuum images demonstrates that extended emission is detected in [\ion{O}{I}]\,$\lambda6300$  but not in \ion{He}{I}\,$\lambda6678$.

 \subsection{A toy model of  \ion{O}{I} emission}
 
 To fit the spatial distribution of [\ion{O}{I}] emission, a toy model is employed  with a power-law radial profile for the [\ion{O}{I}] intensity, $I_{6300}\propto R^{-\alpha}$, where $R$ is the disk radius.  The profile extends from an inner radius $R_{\rm in}$ to an outer radius $R_{\rm out}$, with $I_{6300}$ set to zero when $R<R_{\rm in}$ or  $R>R_{\rm out}$.   The profile is then convolved with the PSF, as measured from nearby continuum regions in the photospheric spectrum. 

A model intensity map is first created from a grid of $\alpha$, $R_{\rm in}$, and $R_{\rm out}$ with the following parameters: $\alpha$ from $1.5$ to $8.5$ in steps of 0.1, $R_{\rm in}$ from 0.06 to 4.22\,AU in steps of 0.04\,AU, and $R_{\rm out}$ from 10 to  82\,AU in steps of 3\,AU.  The model emission is then convolved with the PSF  at the wavelength near the [\ion{O}{I}] line and normalized to the observed [\ion{O}{I}] intensity map using the aperture flux within a radius of 6 pixels from the center. The $\chi^2$ used to judge each model is calculated by comparing the model and observed intensities in a circular region centered on the star with a radius of 55~pixels ($\sim$82.5\,AU).  Only the pixels with S/N$\geq$2 are used for calculating $\chi^2$.

Extended Data Figure~\ref{Fig:reduce_Chi2} shows the reduced $\chi^2$ from the fitting for different sets of $R_{\rm in}$ and $\alpha$ (top-left panel), and $R_{\rm out}$ and $\alpha$ (top-right panel). For each pair of two parameters shown in the figure, the reduced $\chi^2$ is lowest one by setting the third parameter free. The distributions of the reduced $\chi^2$ suggest that   $\alpha$ and $R_{\rm in}$ are constrained very well.  The outer radius $R_{\rm out}$ is not well constrained, with $R_{\rm out}=61$\,AU giving the minimum reduced $\chi^2$ and any value $R_{\rm out}\gtrsim35$\,AU providing a sufficient fit to the observed radial profile of [\ion{O}{I}] emission.  
At those large distances, the [\ion{O}{I}] emission is faint and does not substantially contribute to the total line flux. Given that $R_{\rm out}$ is poorly constrained from fitting, we first fix $R_{\rm out}$=61\,AU, and further refine the grids of $\alpha$ and  $R_{\rm in}$ with $\alpha$ starting from 2.0 to  5.5 with a step of 0.001; $R_{\rm in}$  from 0.06\,AU to  1.54\,AU with a step of 0.01\,AU. The minimum reduced  $\chi^2=1.58$ from the fitting, shown in the bottom panel of Extended Data Figure~\ref{Fig:reduce_Chi2}, has parameters $R_{\rm in}=0.08^{+0.02}_{-0.01}$\,AU  and $\alpha=2.604^{+0.04}_{-0.03}$. Variations of $R_{\rm out}$=41, 51, 71, 81\,AU 
gives $R_{\rm in}=0.07^{+0.01}_{-0.01}$,  $0.07^{+0.01}_{-0.01}$, $0.10^{+0.01}_{-0.01}$, and $0.11^{+0.02}_{-0.01}$, respectively, and $\alpha=2.561^{+0.03}_{-0.03}$, $2.572^{+0.03}_{-0.03}$, $2.652^{+0.03}_{-0.02}$, and $2.677^{+0.03}_{-0.02}$, respectively. The uncertainties correspond to the 68\% confidence level for individual parameters.

\subsection{[O\,{\small I}] and [Ne\,{\small II}] spectral line profiles} With the [\ion{O}{I}]  intensity distribution derived above, we further explore the velocity field of the wind traced by [\ion{O}{I}] emission. The left panel of Extended Data   Figure~\ref{Fig:OIcen} shows the distribution of  the centroids of the [\ion{O}{I}] line profiles derived from the spectroscopic data collected from PolarBase\cite{petit14},  ESO Science Archive Facility, and Keck Observatory Archive. The data obtained with ESPaDOnS were downloaded from PolarBase,  data from FEROS, HARPS, and UVES from the ESO Science Archive Facility, and data from HIRES obtained from the Keck Observatory Archive. These spectra are corrected for the telluric absorption and radial velocities. The telluric absorption models are obtained from TAPAS\cite{bertaux14} for the data taken with the instruments (ESPaDOnS and HIRES) on Mauna~Kea and from the ESO online SKYCALC Sky Model Calculator\cite{noll12} for the data obtained with the ESO instruments (FEROS, HARPS, and UVES). The radial velocities of individual ESPaDOnS, FEROS, HARPS, and UVES spectra spectra are estimated by cross-correlating the observed spectra with synthetic spectrum with $T_{eff}=4000$\,K and logg=3.5 (ref.\cite{husser14}), in the interval from 
6228 and 6272\,\AA, where emission lines are absent and the telluric contamination is negligible. The HIRES spectra are wavelength calibrated using the same technique in the interval from 6277.6 and 6306\,\AA, since the HIRES spectrum used here does not fully cover the 6228-6272\,\AA\ range. The RVs are derived by cross-correlating the individual spectra with the synthetic spectrum, which is veiled, rotationally broadened, and degraded to match the corresponding spectral resolution.

The [\ion{O}{I}]\,$\lambda6300$ line profiles of individual spectra, corrected for the  telluric absorption and radial velocities, are then derived by subtracting best-fit veiled UVES spectra of RECX~10, which has a similar spectral type (M0.5, ref.\cite{fang21}) to TW~Hya, from the corrected spectra of TW~Hya using the same method as done for extracting the [\ion{O}{I}]\,$\lambda6300$ and \ion{He}{I} $\lambda6678$ lines from the MUSE spectra. In the left panel of Extended Data Figure~\ref{Fig:OIcen}, the [\ion{O}{I}] line centroids have a mean of $-$0.8\,\kms\ with a standard deviation of 0.4\,\kms, and 98\% of them are blueshifted, which provides a strong evidence that the [\ion{O}{I}] emission originates in the wind. The right panel of Extended Data Figure~\ref{Fig:OIcen} compares the [\ion{O}{I}] line profiles taken with different instruments. For the comparisons, the line centroids have been shifted to the mean centroid ($-$0.8\,\kms)  and the spectral resolutions have been degraded to 40000. We note that the [\ion{O}{I}] lines show both blue wings and red wings and the red wings  are more variable than the blue wings. However, an investigation of the wing variability is beyond the scope of this work. 

Extended Data Figure~\ref{Fig:lineprofile} shows the combined [\ion{O}{I}]~$\lambda6300$ line profile (left panel) observed with UVES during  29$^{th}$-31$^{th}$ March 2022, and the [\ion{Ne}{II}]~$12.81\,\mu m$ line profile (right panel)
 observed on 23$^{th}$-24$^{th}$ February 2010 (ref.\cite{pascucci11}). The line centroid is blueshifted by 1.8\,\kms\ for [\ion{O}{I}], 1\,\kms\ higher than the mean centroid velocity, and 6~\kms\ for [\ion{Ne}{II}]. The MHD disk wind model that provides the radial emission profile in Figure~\ref{Fig:PEmodel}, also yields spectral line profiles that fit both emission lines on the blue wings fairly well without fine-tuning (Extended Data Figure~\ref{Fig:lineprofile}). The emission of magnetothermal winds is mainly located at low altitudes, where the wind acceleration  
 is far from being finished. Therefore, the blue wing is much slower than the terminal speeds of the winds. The red wing remains unexplained by the MHD wind due to the lack of redshifted emission components in disk winds with a face-on configuration. Our generic model assumes a full disk, and the redshifted emission has been explained by seeing through the inner hole of the disk\cite{pascucci20}.  If winds commonly produce [\ion{O}{I}] emission within 1~AU, as empirically measured here, then double-peaked profiles might be expected for sources viewed edge-on.
An alternate possibility for both the red wing of TW~Hya and the general lack of double-peaked profiles\cite{banzatti19} is that the wind is turbulent enough to broaden the line profile.  However, a turbulent wind is beyond our current simulation model and will be elaborated in future work.

The [\ion{Ne}{II}] line profile\cite{pascucci11} is well reproduced by the generic magnetothermal disk wind model, without any fine tuning to actual disk properties of TW~Hya.  However, the observed flux is three times stronger than the model flux\cite{pascucci09}.  This discrepancy may be explained either by a strong contribution from photoevaporation to the [\ion{Ne}{II}] line flux or by stronger emission in the magnetothermal wind, if the EUV and soft X-ray emission from the star is underestimated.

We limit the current analysis to demonstrating the consistency between the observed line profile and an out-of-the-box model, since models are often flexible enough in tuning to explain any line properties and are therefore less powerful in predictions.  In any case, the [\ion{Ne}{II}] line will need to be spatially resolved in order to demonstrate the mass loss rate in any photoevaporative wind.

\end{methods}

\section{Data availability}
The raw MUSE data can be taken from ESO archive under programme IDs 0104.C-0449. 
The reduced spectral data from other ESO facilities can be download through ESO Phase III archive with programme IDs 074.A-9021(A), 078.A-9059(A), 079.A-9006(A), 079.A-9007(A), 079.A-9017(A), 081.A-9005(A), 081.A-9023(A), 089.A-9007(A), 0101.A-9012(A), 0102.A-9008(A), 60.A-9036(A), 075.C-0202(A), 081.C-0779(A), 082.C-0390(A), 082.C-0427(C), 65.I-0404(A), 082.C-0218(A), 089.C-0299(A), 106.20Z8.011.  The reduced Keck spectral data are download from Keck Observatory Archive  with programme IDs C179Hr, C269Hr, C199Hb,   N125Hr   C186Hr, N204Hr, C189Hr, C252Hr, C247Hr. The reduced ESPaDOnS spectra are downloaded from PolarBase.

\section{Code availability}
The MUSE data are reduced with the MUSE pipeline v2.8.5 which is publicly available. Upon request, the first author will provide code (IDL) to generate figures.

\begin{addendum}
\item We acknowledge the science research grants from the China Manned Space Project with NO. CMS-CSST-2021-B06. This research is based on observations made with ESO Telescopes under programme IDs 0104.C-0449, 074.A-9021(A), 078.A-9059(A), 079.A-9006(A), 079.A-9007(A), 079.A-9017(A), 081.A-9005(A), 081.A-9023(A), 089.A-9007(A), 0101.A-9012(A), 0102.A-9008(A), 60.A-9036(A), 075.C-0202(A), 081.C-0779(A), 082.C-0390(A), 082.C-0427(C), 65.I-0404(A), 082.C-0218(A), 089.C-0299(A), 106.20Z8.011.
 This research has made use of the Keck Observatory Archive (KOA), which is operated by the W. M. Keck Observatory and the NASA Exoplanet Science Institute (NExScI), under contract with the National Aeronautics and Space Administration. We acknowledge S. E. Dahm, L. Hillenbrand, J. Carpenter, W. Borucki, G. W. Marcy for datasets that were been obtained through KOA. 
  \item[Author contributions] 
  G. J. H started the project. J. H reduced the data. M. F. analyzed the data. L. W. contributed in the MHD modeling. A. N. and L. W. performed the line synthesis. M. F., G. J. H, and L. W. were the primary writers of the manuscript. All the authors contributed to   the scientific interpretation of the results and reviewed the manuscript. 
  \item[Competing interests] The authors declare that they have no
competing financial interests.
\item[Additional information]Correspondence and requests for materials should be addressed to M.F. (Email: mfang@pmo.ac.cn).
\end{addendum}


\clearpage

\setcounter{figure}{0}   
\begin{figure}
\centering
\includegraphics[width=3.2in]{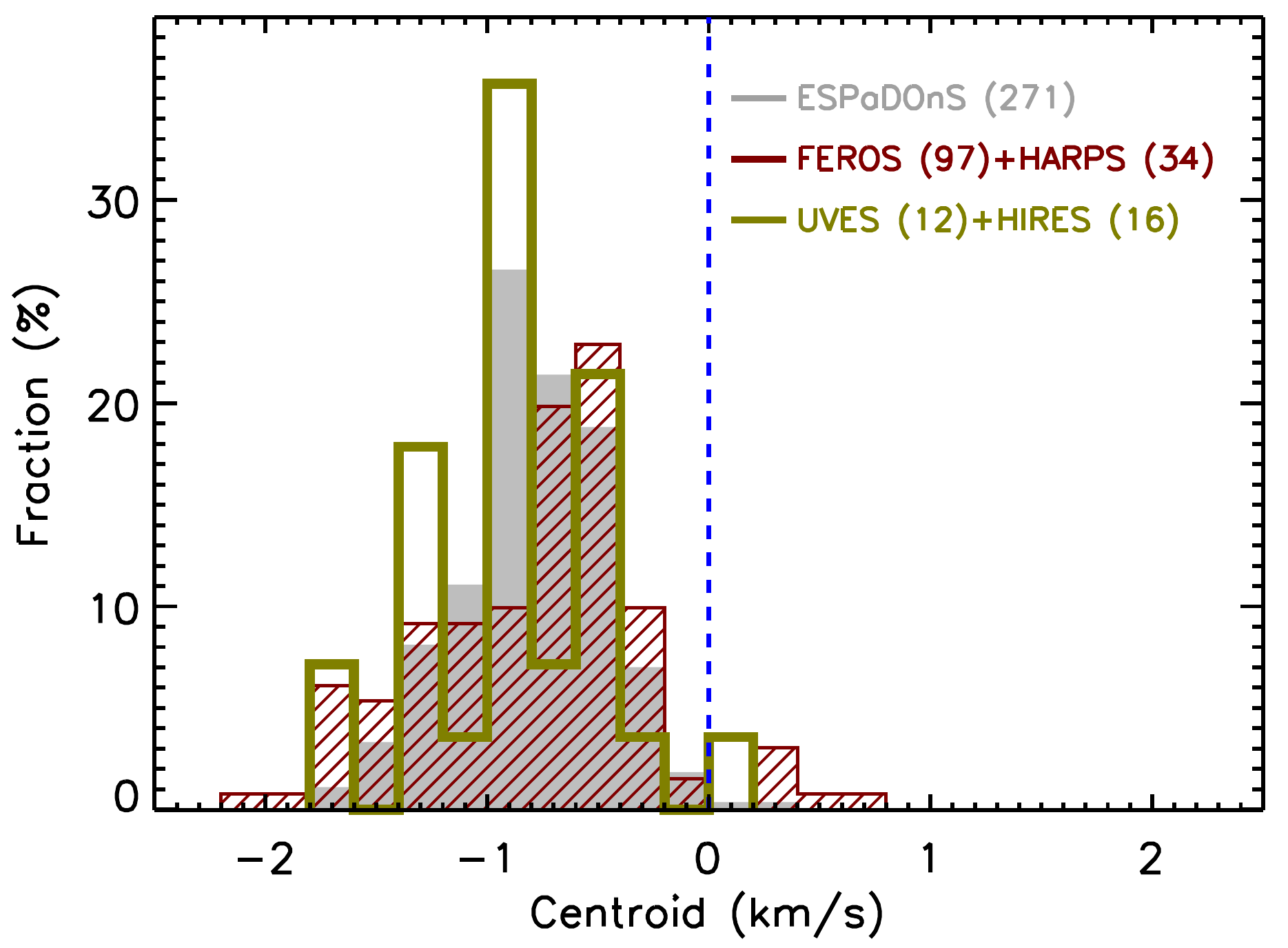}
\includegraphics[width=3.2in]{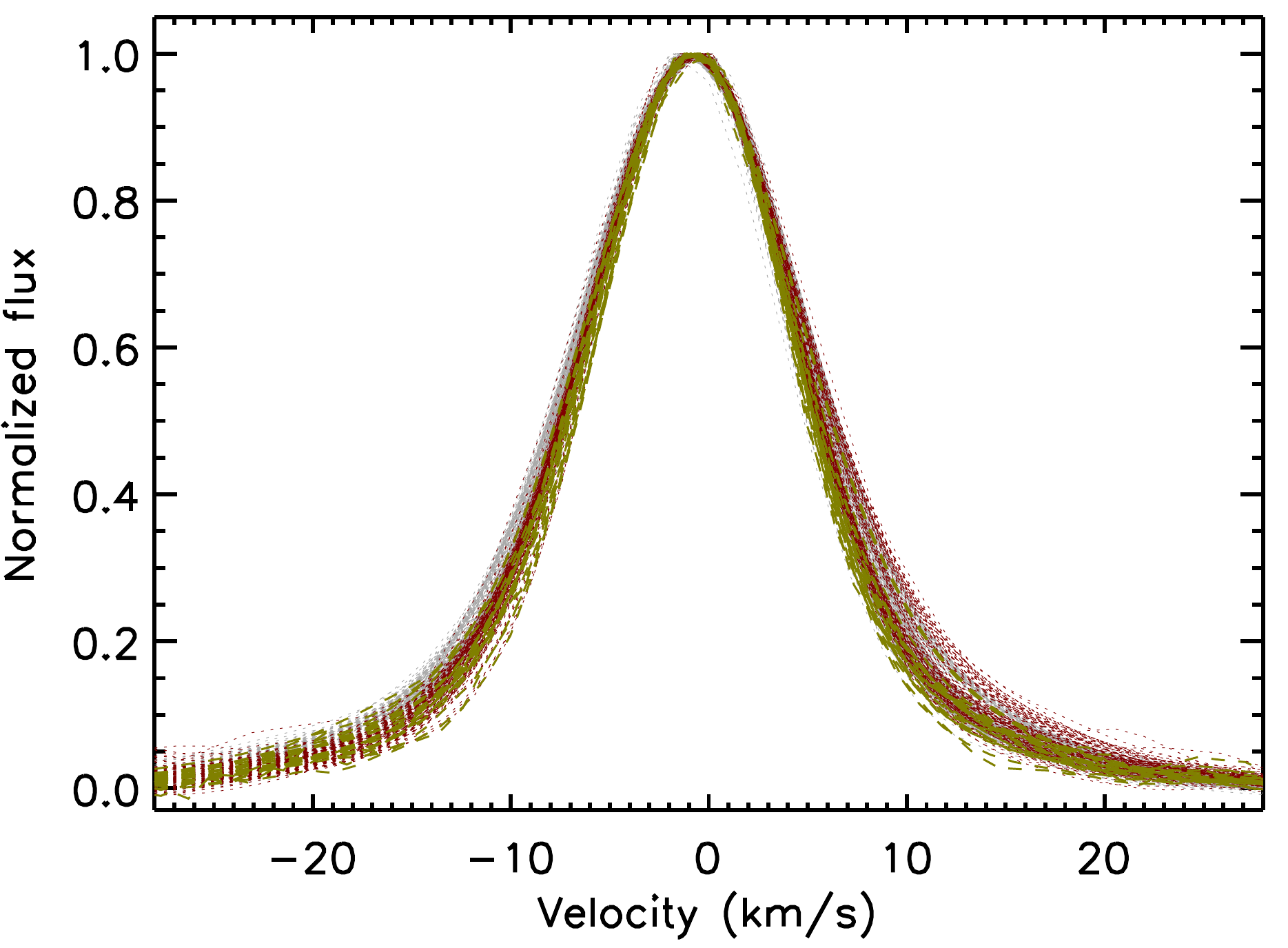}
\linespread{1.0}\selectfont{}
\renewcommand{\figurename}{Extended Data Figure}
\caption{\textbf{[\ion{O}{I}]~$\lambda6300$ line profiles of TW~Hya.} Left Panel: Distribution of the [\ion{O}{I}]~$\lambda6300$ line centroids of TW~Hya and. These spectra are taken with ESPaDOnS (gray color filled histogram), FEROS or HARPS (maroon line filled histogram), and UVES or HIRES (olive open histogram).  The numbers of the used spectra from individual instruments are shown in the figure. Right panel: Comparisons of the [\ion{O}{I}]~$\lambda6300$ line profiles of TW~Hya. For the comparisons, the centroids of the lines have been shifted to the mean centroid ($-$0.8\,\kms) and the spectral resolutions have been degraded to 40000. The colors are the same as in the left panel. }
\label{Fig:OIcen}
\end{figure}

\begin{figure}
\centering
\includegraphics[width=3.2in]{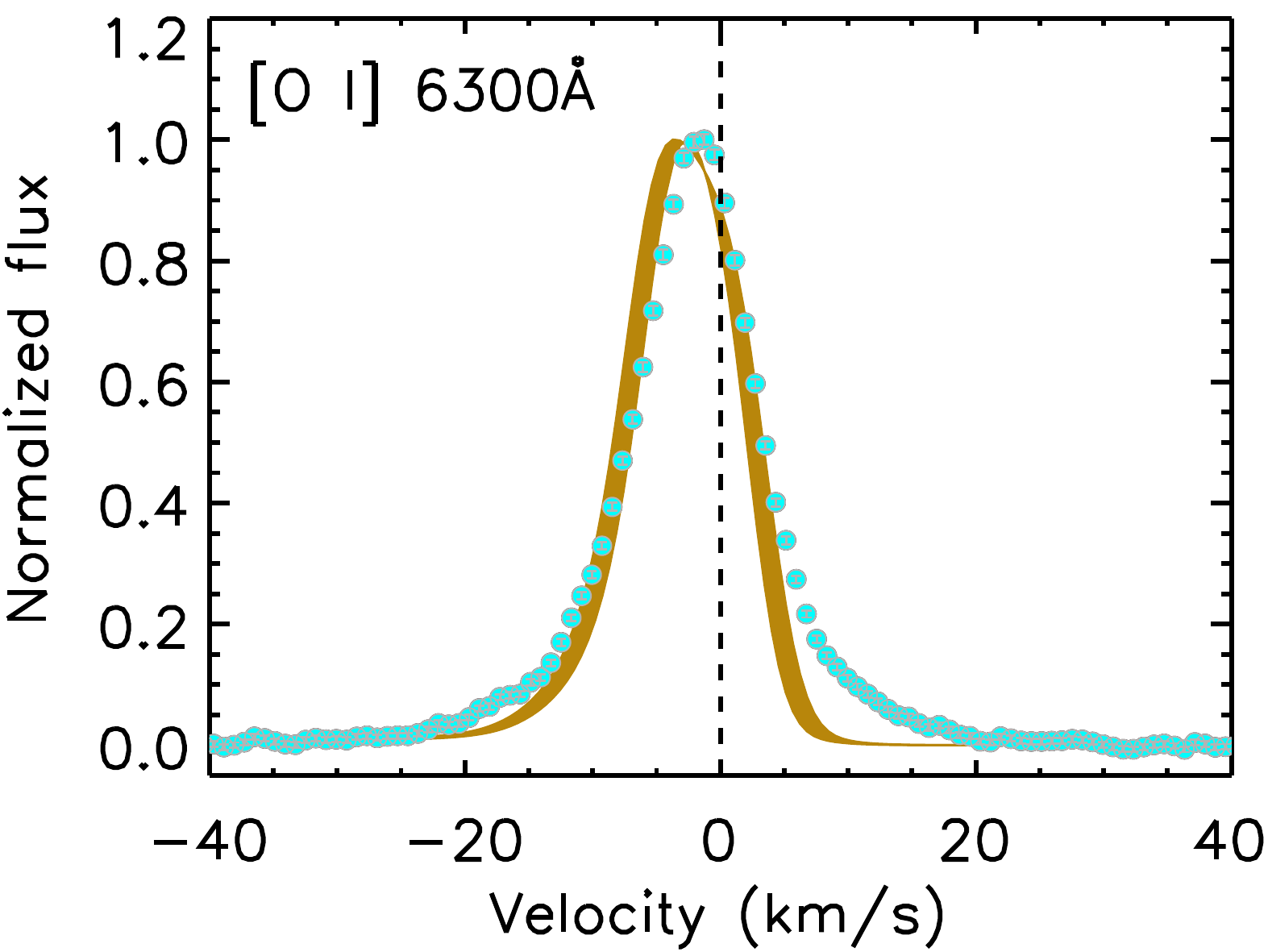}
\includegraphics[width=3.2in]{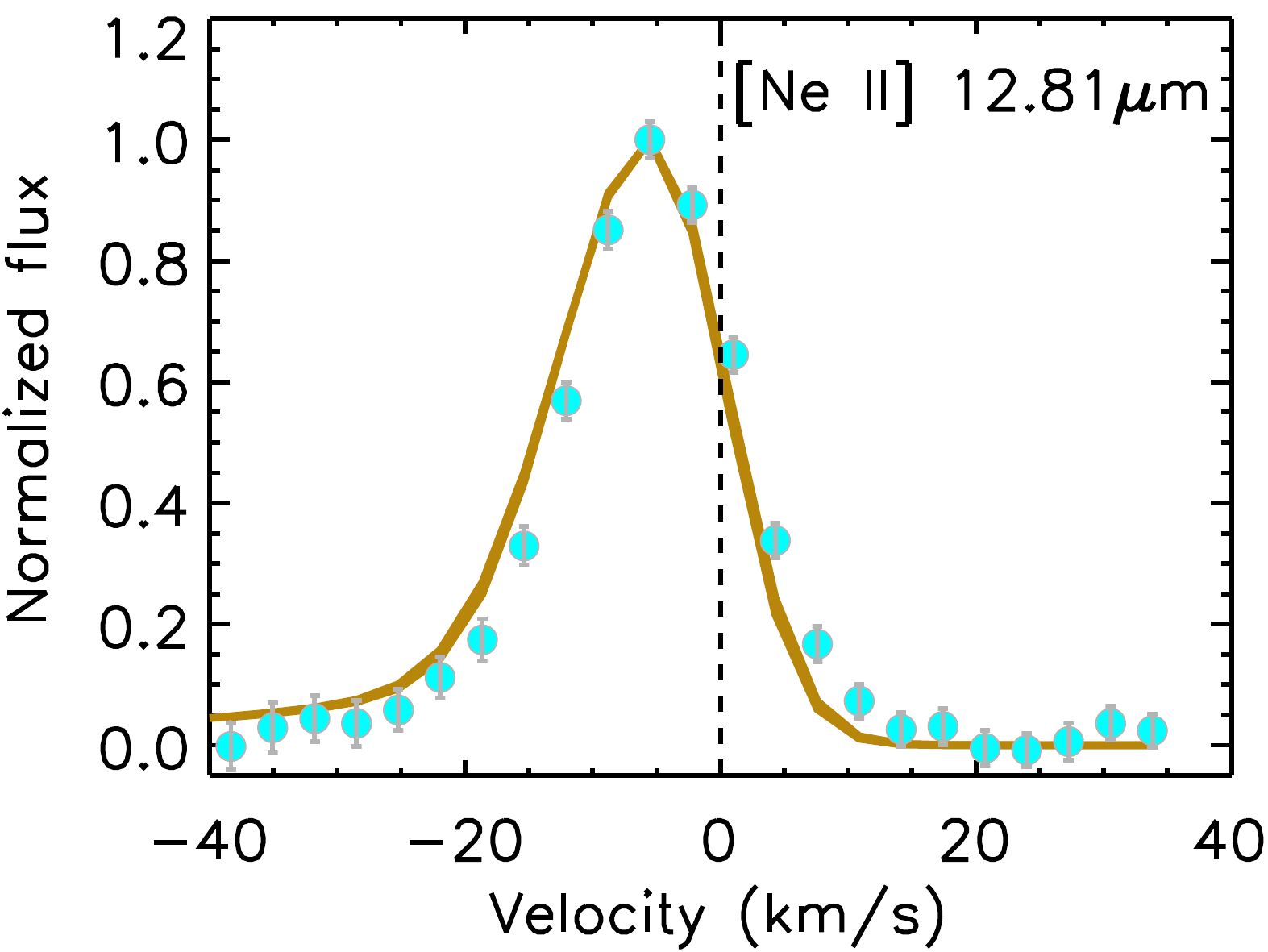}
\linespread{1.0}\selectfont{}
\renewcommand{\figurename}{Extended Data Figure}
\caption{\textbf{[\ion{O}{I}]~$\lambda6300$ and [\ion{Ne}{II}]~$12.81\,\mu m$ line profiles.} Comparisons of the  observed [\ion{O}{I}]~$\lambda6300$ (left) and [\ion{Ne}{II}]~12.81\,$\mu m$ (right)  line profiles  (cyan filled circles) with the simulated ones from the MHD wind model (golden-color filled area for disk inclinations ranging from 5$^{\circ}$ to 7$^{\circ}$).}
\label{Fig:lineprofile}
\end{figure}

\begin{figure}
\includegraphics[width=6.5in]{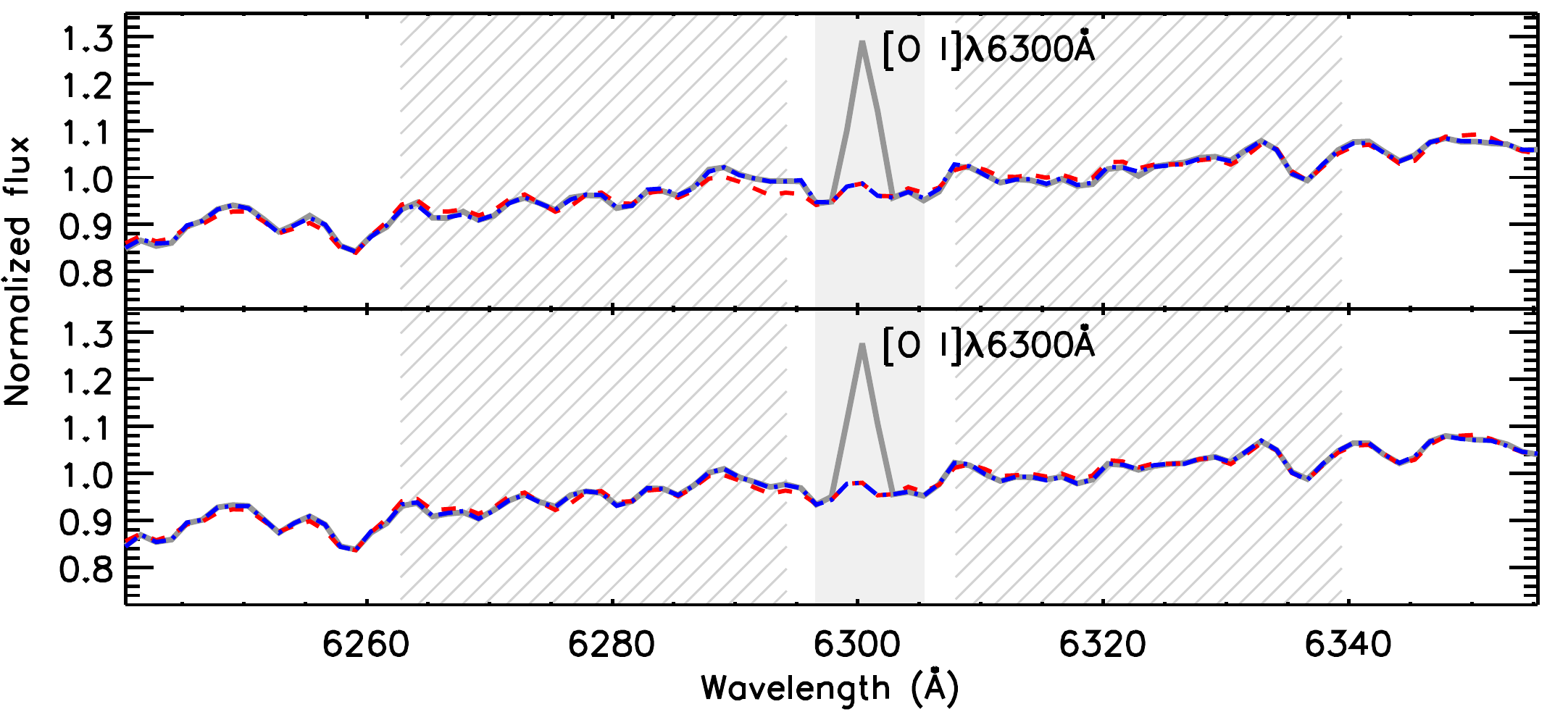}
\includegraphics[width=6.5in]{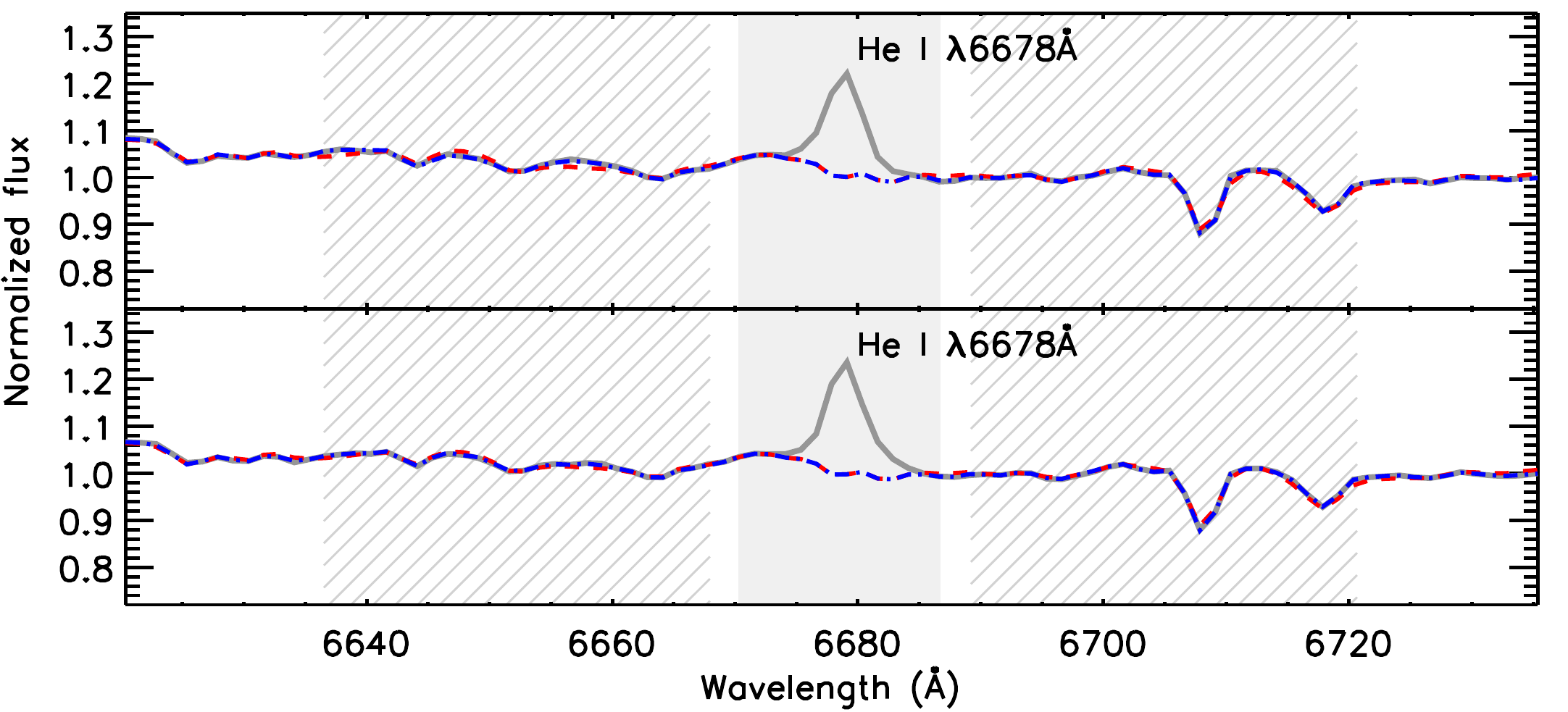}
\linespread{1.0}\selectfont{}
\renewcommand{\figurename}{Extended Data Figure}
\caption{\textbf{Template fits to the observed MUSE spectra near [\ion{O}{I}]~$\lambda6300$ and \ion{He}{I}~$\lambda6678$ line.} Top two panels show the example spectra near [\ion{O}{I}]~$\lambda6300$ (gray lines) of two exposures. In each panel, the red dashed line is the best-fit template, and the blue dash-dotted line for the best-fit template correcting for the residual. The grey color filled area marks the wavelength range within which the [\ion{O}{I}]~$\lambda6300$ line is extracted, and the grey line-filed areas mark the wavelength ranges within which the uncertainties are estimated and the fluxes are integrated to construct PSF. Bottom two panels are same as the top two panels, but for \ion{He}{I}~$\lambda6678$ line.}
\label{fig:example_line_OI6300_HeI6678}
\end{figure}

\begin{figure}
\includegraphics[width=3.25in]{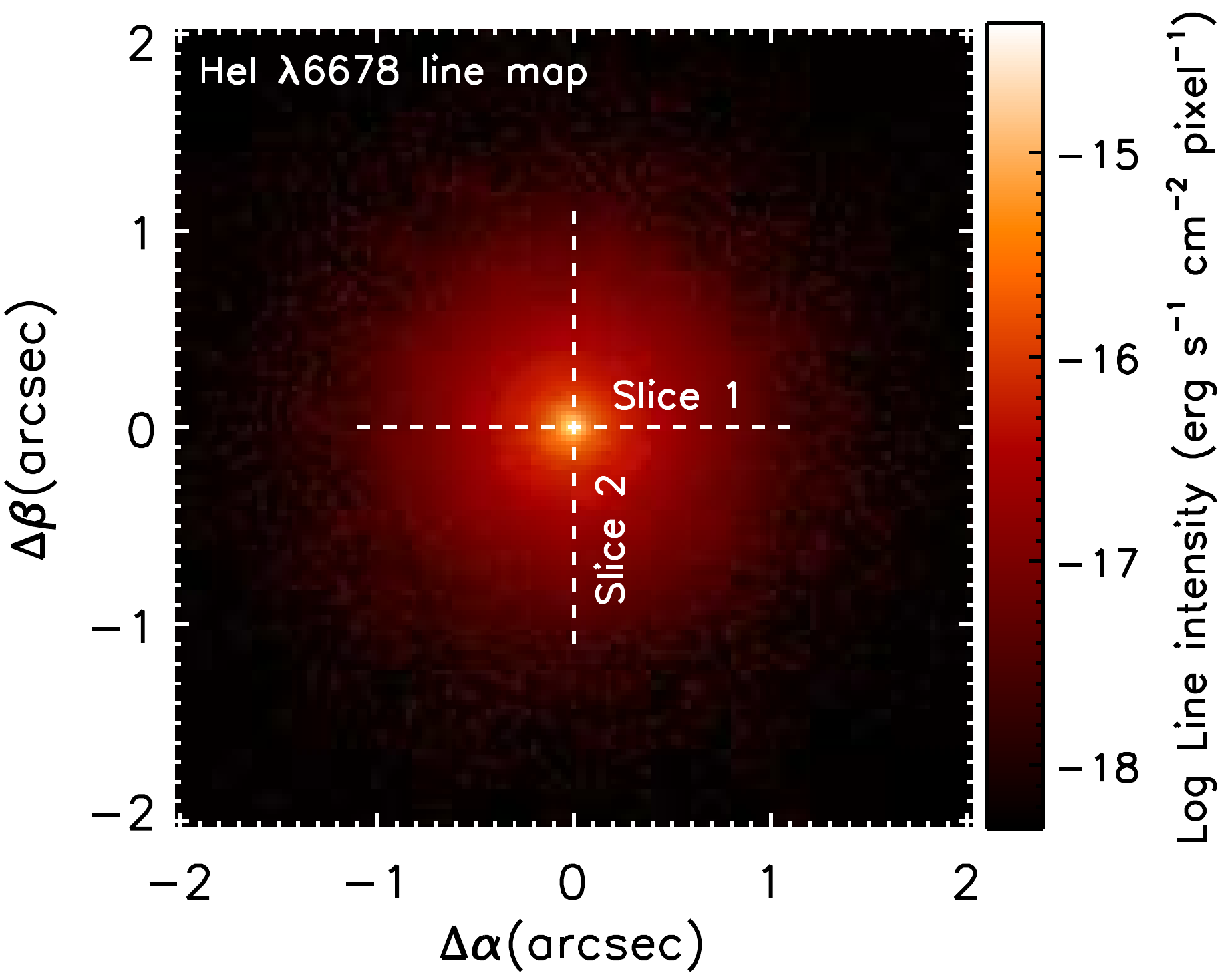}
\includegraphics[width=3.25in]{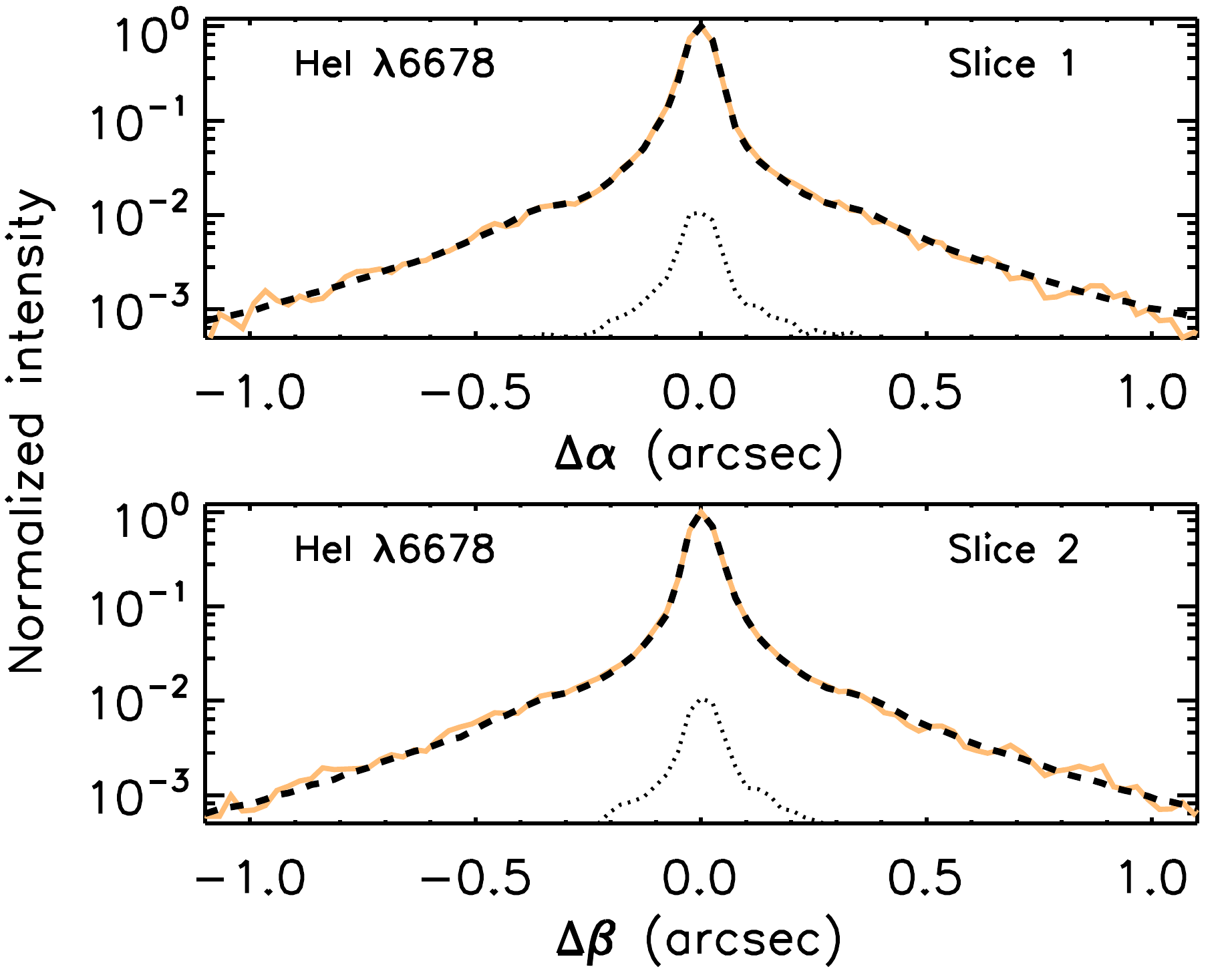}
\includegraphics[width=3.25in]{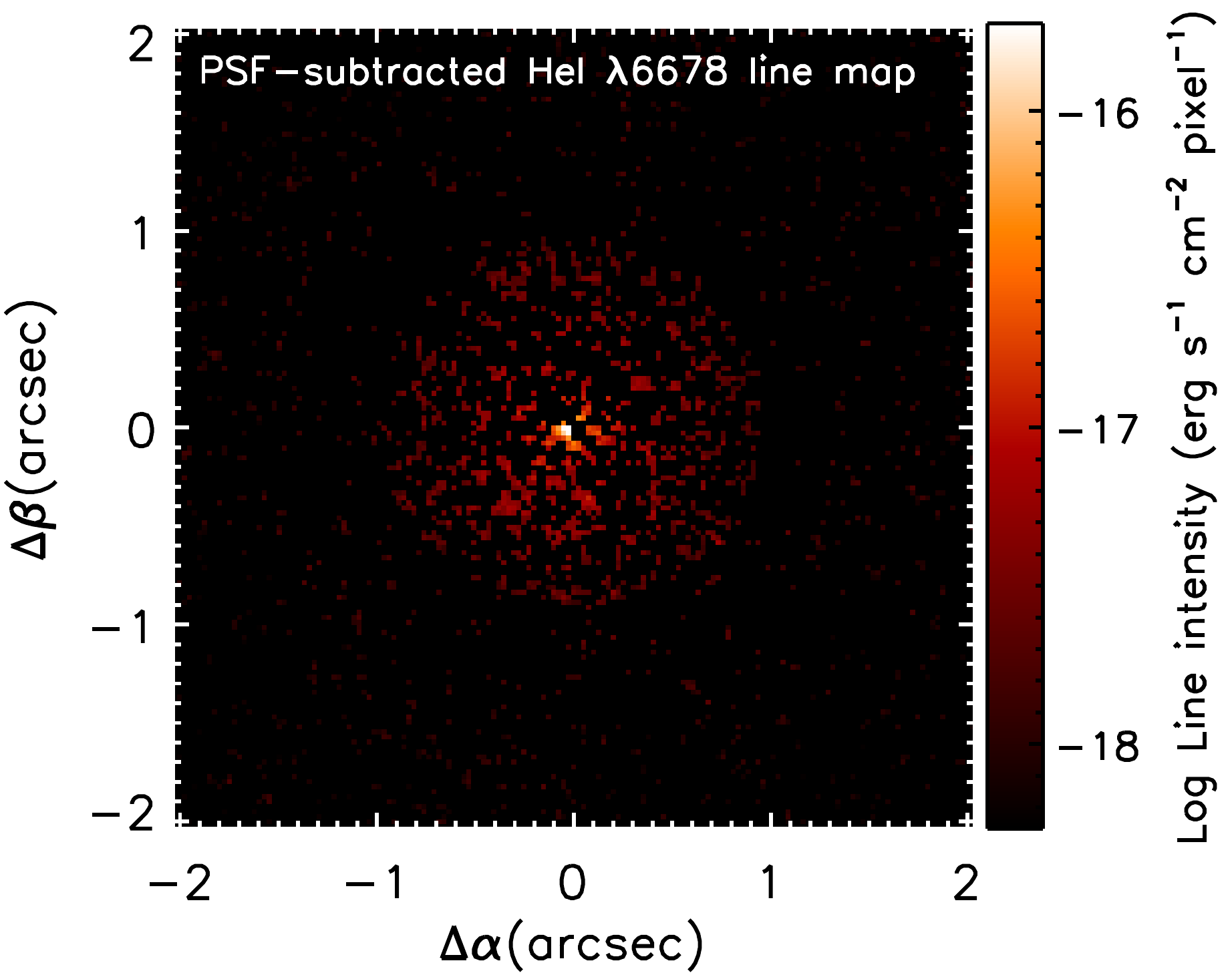}
\includegraphics[width=3.25in]{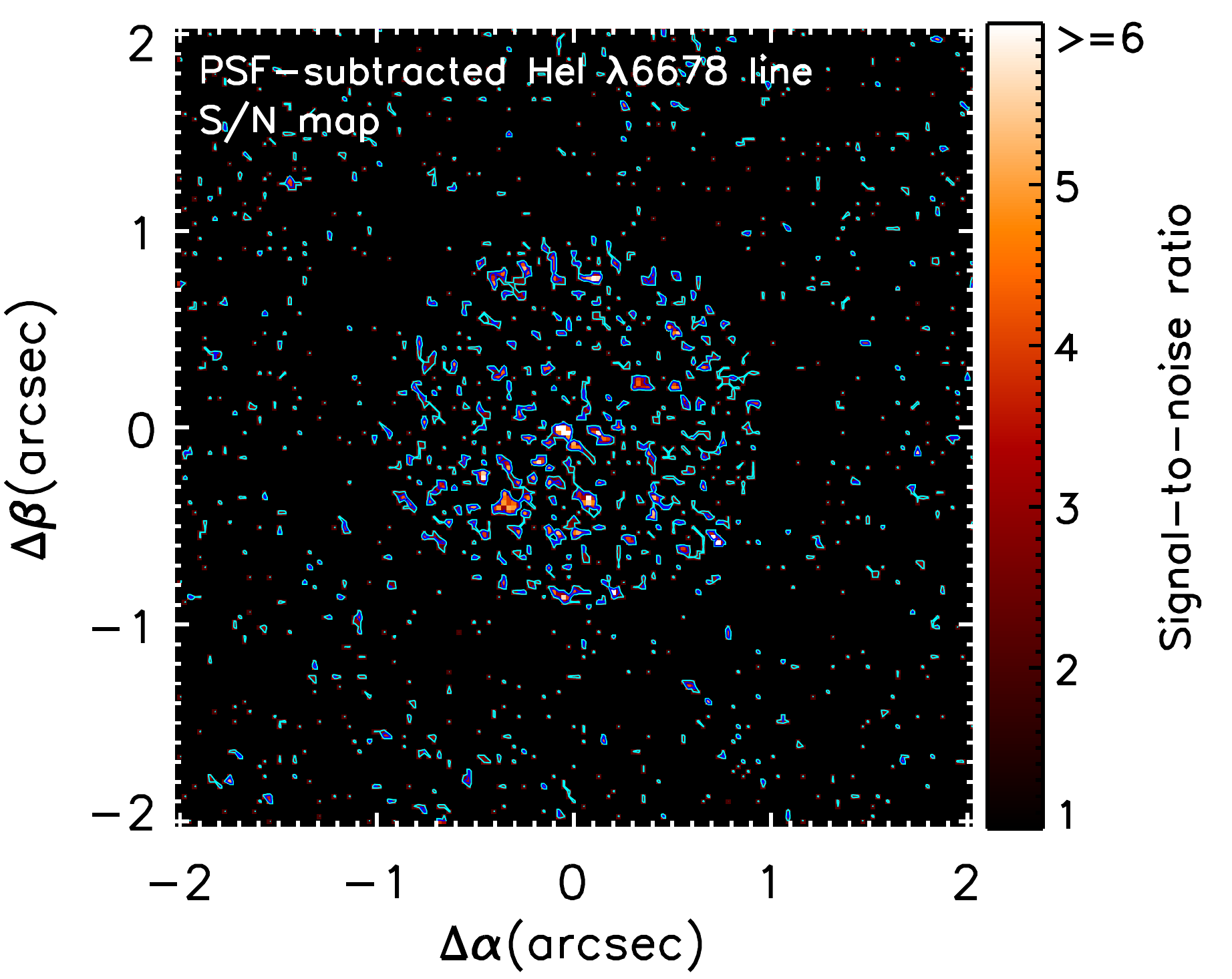}
\linespread{1.0}\selectfont{}
\renewcommand{\figurename}{Extended Data Figure}
\caption{\textbf{Error-weighted mean \ion{He}{I} $\lambda6678$ intensity map.} Upper-left panel: Error-weighted mean \ion{He}{I} $\lambda6678$ intensity map of TW~Hya. Upper-right panels: Two slice cuts (yellow lines) for  \ion{He}{I} $\lambda6678$ line map shown in the upper-left panel, overplotted with the PSF (dark dashed lines) near \ion{He}{I} $\lambda6678$. The dotted line in each panel is the standard deviation along the slice cut. Lower-left panel: \ion{He}{I} $\lambda6678$ line residual map after a subtraction of a PSF which has been normalized to line map by the peak emission.  Lower-right panel: the signal-to-noise ratio map for the \ion{He}{I} $\lambda6678$ line residual map shown in the Lower-left panel. The cyan and blue lines show contours with S/N=2 and 3, respectively.
}
\label{Fig:HeI6678_mean}
\end{figure}

\begin{figure}
\centering
\includegraphics[width=3.22in]{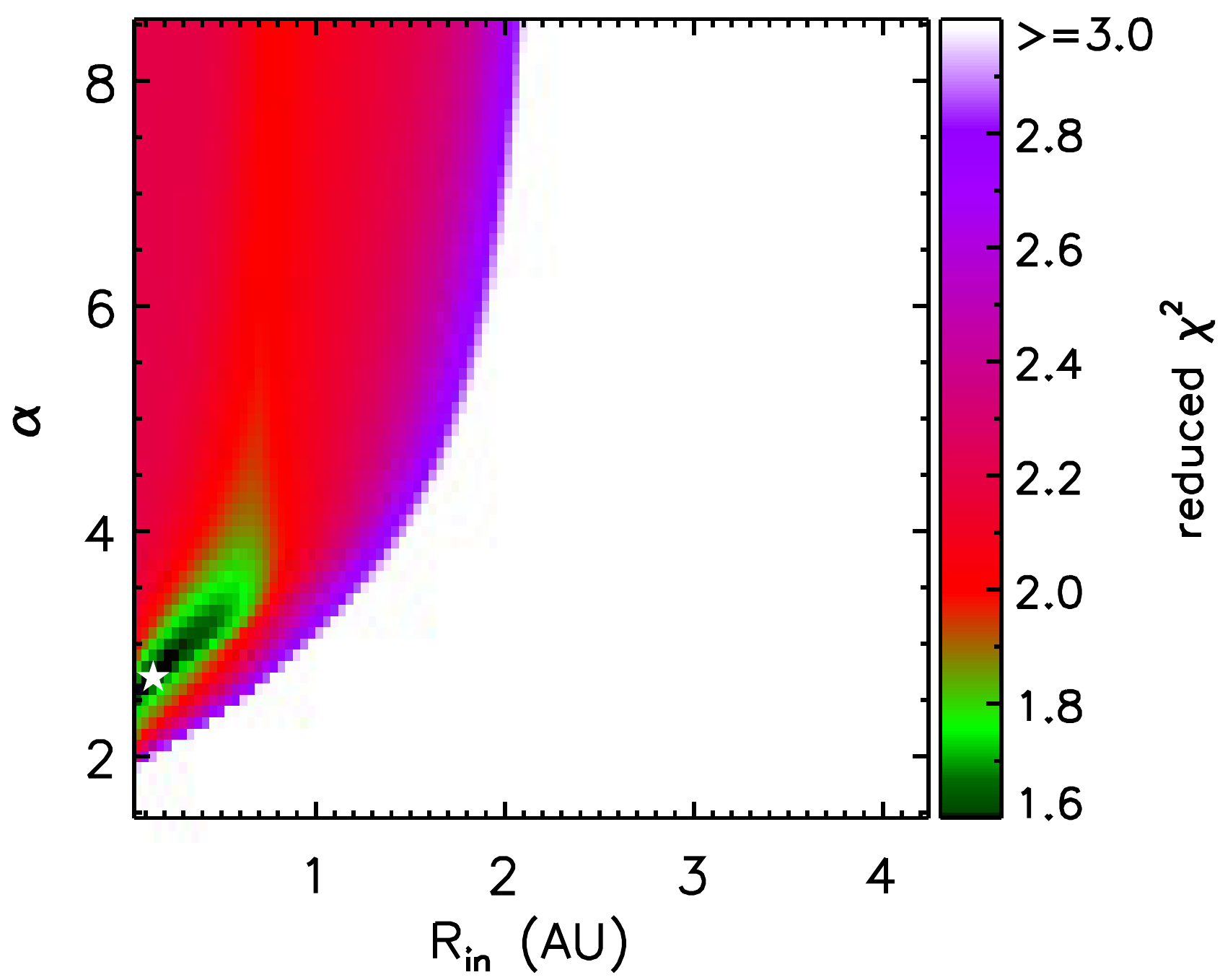}
\includegraphics[width=3.22in]{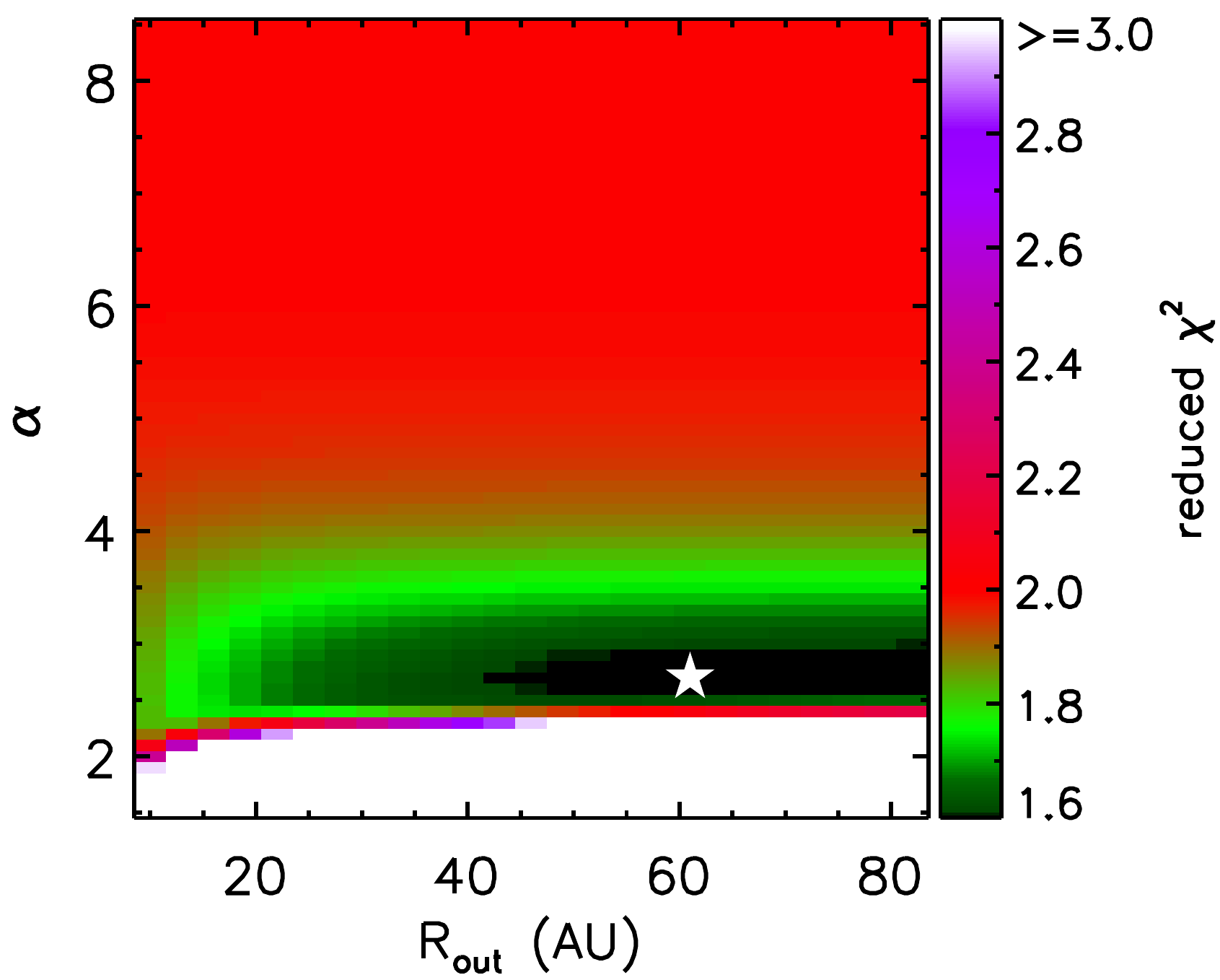}
\includegraphics[width=3.22in]{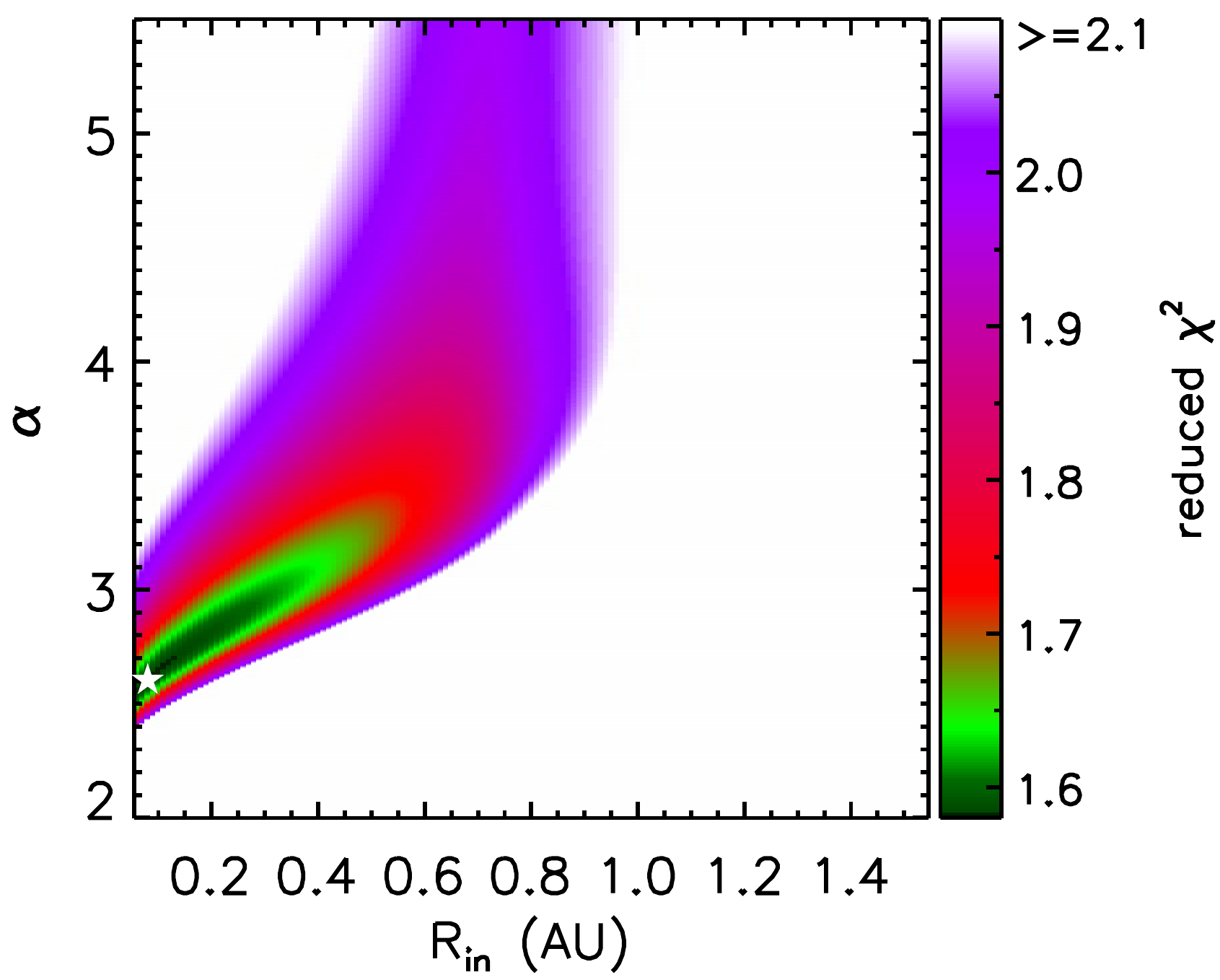}
\linespread{1.0}\selectfont{}
\renewcommand{\figurename}{Extended Data Figure}
\caption{\textbf{Reduced $\chi^2$ for the toy models.} Top two panels: Reduced $\chi^2$ from the coarse grids of the three free parameters:  $\alpha$, $R_{\rm in}$ and $R_{\rm out}$. In each panel, the reduced $\chi^2$ for each pair of two parameters is lowest one by setting the remaining parameter free. Bottom panel: Reduced $\chi^2$ from the fine grid of $\alpha$ and $R_{\rm in}$ with $R_{\rm out}=61$\,AU. In each panel, the pentagram symbol marks the minimum reduced $\chi^2$.}
\label{Fig:reduce_Chi2}
\end{figure}





\end{document}